\documentclass[apj]{emulateapj}
\usepackage{rotating}
\usepackage{lscape}

\newcommand{\IUE}{{\it IUE}}
\newcommand{\HST}{{\it HST}}
\newcommand{\kms}{\ifmmode {\rm km\ s}^{-1} \else km s$^{-1}$\fi}
\newcommand{\Msun}{\ifmmode {\rm M}_{\odot} \else M$_{\odot}$\fi}
\newcommand{\Lsun}{\ifmmode {\rm L}_{\odot} \else L$_{\odot}$\fi}
\newcommand{\qo}{\ifmmode q_{\rm o} \else $q_{\rm o}$\fi}
\newcommand{\Ho}{\ifmmode H_{\rm o} \else $H_{\rm o}$\fi}
\newcommand{\ho}{\ifmmode h_{\rm o} \else $h_{\rm o}$\fi}
\newcommand{\ltsim}{\raisebox{-.5ex}{$\;\stackrel{<}{\sim}\;$}}

\newcommand{\vFWHM}{\ifmmode v_{\mbox{\tiny FWHM}} \else
                    $v_{\mbox{\tiny FWHM}}$\fi}
\newcommand{\CCF}{\ifmmode F_{\it CCF} \else $F_{\it CCF}$\fi}
\newcommand{\ACF}{\ifmmode F_{\it ACF} \else $F_{\it ACF}$\fi}
\newcommand{\Halpha}{\ifmmode {\rm H}\alpha \else H$\alpha$\fi}
\newcommand{\Hbeta}{\ifmmode {\rm H}\beta \else H$\beta$\fi}
\newcommand{\Hgamma}{\ifmmode {\rm H}\gamma \else H$\gamma$\fi}
\newcommand{\Hdelta}{\ifmmode {\rm H}\delta \else H$\delta$\fi}
\newcommand{\Lya}{\ifmmode {\rm Ly}\alpha \else Ly$\alpha$\fi}
\newcommand{\Lyb}{\ifmmode {\rm Ly}\beta \else Ly$\beta$\fi}
\newcommand{\HeI}{\ifmmode {\rm He}\,{\sc i}\,\lambda5876 \else 
	          He\,{\sc i}\,$\lambda5876$\fi}
\newcommand{\HeII}{\ifmmode {\rm He}\,{\sc ii}\,\lambda4686 \else 
	           He\,{\sc ii}\,$\lambda4686$\fi}

\newcommand{\heii}{He\,{\sc ii}}

\newcommand{\ciii}{\ifmmode {\rm C}\,{\sc iii} \else C\,{\sc iii}\fi}
\newcommand{\civ}{\ifmmode {\rm C}\,{\sc iv} \else C\,{\sc iv}\fi}
\newcommand{\CIV}{\ifmmode {\rm C}\,{\sc iv}\,\lambda1549 \else 
	           C\,{\sc iv}\,$\lambda1549$\fi}

\newcommand{\niv}{N\,{\sc iv}}

\newcommand{\oiii}{O\,{\sc iii}}
\newcommand{\ob}{[O\,{\sc iii}]\,$\lambda \lambda 4959,5007$}

\newcommand{\mgii}{Mg\,{\sc ii}}
\newcommand{\MGII}{Mg\,{\sc ii}\,$\lambda2800$}

\newcommand{\civm}{{\rm C}\,{\scriptstyle{\rm IV}}}

\shorttitle{CIV Line Absorption}
\shortauthors{}

\accepted{7 August 2013}

\begin{document}

\title{C\,{\small{IV}} Line-Width Anomalies: The Perils of Low $S/N$
  Spectra}

\author{ K.~D.~Denney\altaffilmark{1,6}, R.~W.~Pogge\altaffilmark{2},
  R.~J.~Assef\altaffilmark{3,7}, C.~S.~Kochanek\altaffilmark{2,4},
  B.~M.~Peterson\altaffilmark{2,4}, M.~Vestergaard\altaffilmark{1,5}}

\altaffiltext{1}{Dark Cosmology Centre, 
                 Niels Bohr Institute, 
                 Juliane Maries Vej 30, 2100 Copenhagen \O, Denmark;
                 kelly@dark-cosmology.dk}

\altaffiltext{2}{Department of Astronomy, 
		The Ohio State University, 
		140 West 18th Avenue, 
		Columbus, OH 43210, USA}

\altaffiltext{3}{Jet Propulsion Laboratory, 
                 California Institute of Technology, 
                 MS 169-530, 4800 Oak Grove Drive, Pasadena, 91109, USA}

\altaffiltext{4}{Center for Cosmology and AstroParticle Physics, 
                 The Ohio State University,
		 191 West Woodruff Avenue, 
		 Columbus, OH 43210, USA}

\altaffiltext{5}{Steward Observatory and Department of Astronomy, 
                 University of Arizona, 
                 933 N Cherry Ave, 
                 Tucson, AZ 85721, USA}



\footnotetext[6]{Marie Curie Fellow}
\footnotetext[7]{NASA Postdoctoral Fellow}

\begin{abstract}
  Comparison of six high-redshift quasar spectra obtained with the Large
  Binocular Telescope with previous observations from the Sloan Digital
  Sky Survey shows that failure to correctly identify absorption and
  other problems with accurate characterization of the \CIV\ emission
  line profile in low $S/N$ data can severely limit the reliability of
  single-epoch mass estimates based on the \civ\ emission line.  We
  combine the analysis of these new high-quality data with a reanalysis
  of three other samples based on high $S/N$ spectra of the \civ\
  emission line region.  We find that a large scatter between the
  \Hbeta- and \civ-based masses remains even for this high $S/N$ sample
  when using the FWHM to characterize the BLR velocity dispersion and
  the standard virial assumption to calculate the mass.  However, we
  demonstrate that using high-quality data and the line dispersion to
  characterize the \civ\ line width leads to a high level of consistency
  between \civ- and \Hbeta-based masses, with $<0.3$\, dex of observed
  scatter, and an estimated $\sim$0.2\, dex intrinsic scatter, in the
  mass residuals.
\end{abstract}



\section{INTRODUCTION}

In the well-accepted paradigm of galaxy formation and evolution from
hierarchical structure growth, it is simple enough to reach the
conclusion that all massive galaxies should house supermassive black
holes (BHs; see also \citealt{Soltan82}).  Unfortunately, it is not as
simple a task to measure BH masses, growth rates, and their evolution.
BH mass measurements using dynamical methods in quiescent galaxies (see
e.g., \citealt{Gebhardt03} and compilations such as
\citealt{Ferrarese05, Graham08, Gultekin09}) require high spatial
resolution and are thus restricted to the local universe.  On the other
hand, reverberation mapping \citep[RM;][]{Blandford82, Peterson93} is a
successful method for directly measuring BH masses in active galaxies.
While not restricted by spatial resolution, reverberation mapping is
dependent on temporal resolution.  Thus, distance is not a fundamental
restriction, but obtaining the long-term observing resources to meet the
temporal sampling requirements has logistically driven most
reverberation experiments to target lower-luminosity, faster varying
AGNs in the local universe, $z\lesssim 0.3$.  Nonetheless, scaling
relationships derived from reverberation mapping results of the local
AGN population \citep[e.g.][]{Vestergaard06, McGill08, Rafiee11,
  Bentz13} enable a means for studying BH masses in Type 1 (broad-line)
AGNs at all redshifts based on single-epoch (SE) mass estimates
\citep[see, e.g.,][]{Vestergaard04, Vestergaard&Osmer09, Kelly10,
  Shen11}.

These SE BH mass estimates depend on several assumptions.  First, the
luminosity of the AGN continuum, measured from the spectrum near a broad
emission line of interest, must be a valid proxy for the broad line
region radius (BLR).  In theory, this is expected because
photoionization physics regulates the production of line-emitting
photons in such a way that the characteristic radius of emission,
$R_{\rm BLR}$, scales tightly with the nuclear luminosity, $L$
\citep{Davidson72, Krolik&McKee78}.  More importantly, direct
reverberation mapping measurements of $R_{\rm BLR}$ show the expected
correlation and provide an empirically well-calibrated $R_{\rm BLR}-L$
relation \citep{Kaspi00, Bentz09rl, Bentz13}.  The indirect BLR radius
is then combined with a broad emission-line width from at least one
broad emission line (\Halpha, \Hbeta, \MGII, or \CIV) with a calibrated
scaling relation.  This width is assumed to be representative of the BLR
gas velocity under the influence of the gravity of the central black
hole.  Given these assumptions, the virial BH mass is estimated by
$M_{\rm BH}=fR_{\rm BLR}(\Delta V)^2/G$, where $R_{\rm BLR}$ scales as
$\sim L^{1/2}$, $\Delta V$ is the velocity dispersion of the BLR gas,
$G$ is the gravitational constant, and $f$ is a dimensionless factor of
order unity accounting for the unknown BLR geometry and kinematics and
determined from local calibrations \citep[cf.][]{Onken04, Woo10,
  Park12b, Grier13b}.

At redshifts $z \gtrsim 2$, all emission lines but \CIV\ have redshifted
out of the optical observing window, making it the only emission line
available for high-$z$ BH and galaxy evolution studies using
ground-based optical data.  However, several past studies have claimed
that \civ\ is an unreliable virial mass indicator (e.g.,
\citealt{Baskin&Laor05}; \citealt{Sulentic07}; \citealt{Netzer07},
hereafter N07; \citealt{Shen12}; \citealt{Benny&Netzer12}) due to large
scatter and possible offsets in the \civ\ based masses compared to
\Halpha, \Hbeta, or \mgii.  The most wide-spread, physically motivated
argument against \civ\ is tied to the commonly observed blueward
asymmetries, enhancements, and velocity shifts of the \civ\ line
profile.  It has been suggested that these observed properties are the
result of non-virial motions of the \civ-emitting gas \citep[i.e.,
outflows, winds, and non-gravitational forces;][]{Gaskell82, Wilkes84,
  Richards02, Leighly04a}, rendering \civ\ velocity width measurements
unsuitable for estimating BH masses.  We should note, however, that as
with stellar winds, any radiatively driven wind will result in a
velocity comparable to the escape velocity \citep{Cassinelli73}, which
is close enough to the virial velocity that it is unlikely to be a
considerable issue given the other uncertainties in the problem.  Other
studies have found general consistency between single-epoch \civ\ and
\Hbeta\ masses (e.g., \citealt{Vestergaard06}, hereafter VP06;
\citealt{Greene10}; \citealt{Assef11}, hereafter A11), suggesting that
any biases are modest.  The only way to definitively probe the \civ\ BLR
kinematics and search for potential non-virial motions is using
reverberation mapping experiments of the \civ\ emission.  These
experiments isolate the photoionized, and apparently virialized
\citep{Peterson99b, Peterson00a}, gas in the BLR that is responding to
the continuum variability from other non-variable emission components.
Further constraints on the geometry and kinematics are then possible
with two-dimensional velocity--delay maps \citep[see, e.g.,][]{Horne04,
  Bentz10b, Pancoast12, Grier13}, but unambiguous maps have yet to be
constructed for this emission line \citep[though see][]{Ulrich96}.
Nonetheless, available reverberation mapping results for \civ\ yield
consistent results with those of the other emission lines observed in
the same objects \citep{Peterson99b, Peterson00a, Peterson04},
so any issues are restricted to SE estimates using \civ, rather than
\civ\ in general.

One concern contributing to the \civ\ debate is that sample selection
may be a problem.  For example, VP06 studied only local
reverberation-mapped AGNs.  \citet{Richards11} show that this sample
does not span the full \civ\ equivalent width/blueshift parameter space
observed for Sloan Digital Sky Survey (SDSS) quasars, suggesting that
the results may not be representative of the overall high-redshift
quasar population, and raise the concern that BH mass scaling
relationships calibrated only with low blueshift sources may not be
applicable to the large blueshift, low equivalent width
(``wind-dominated'') sources. On the other hand, the VP06 reverberation
mapping sample spans a rest-UV luminosity range of 3 orders of
magnitude.  This is much larger than most studies (e.g., N07;
\citealt{Dietrich09}, hereafter D09; \citealt{Shen12}; \citealt{Ho12}),
making it far easier to recognize the existence of an underlying
correlation in the presence of noise.  Indeed, the studies finding
little or no correlation (e.g., D09; \citealt{Greene10};
\citealt{Shen12}; \citealt{Ho12}) first restrict the sample to such a
narrow luminosity range that no correlation would be found for any
estimator, including \Hbeta.  A11 pointed out that roughly half of the
`problem' has nothing to do with the line widths but comes from the
variance between the continuum estimates rather than the line structure.
\citet{Denney12} then argued that much of the discrepancy due to the
line widths between \Hbeta\ and \civ-based SE BH mass estimates is due
to a non-variable component of the \civ\ emission-line that biases the
full width at half maximum (FWHM) line widths often used to derive the
SE BH mass.  Since the component biasing the FWHM seems not to
reverberate, direct BH mass measurements based on reverberation mapping
are unaffected, leading to the better agreement with results for \Hbeta.

\civ\ supporters also argue that data quality is a key factor: VP06
largely used high-$S/N$ {\it HST} spectra or an average reverberation
mapping campaign spectrum, and A11 obtained new or previously published
high-$S/N$ \civ\ spectra of all their targets.  Most studies instead use
\civ\ observations in lower $S/N$ survey spectra, such as from SDSS.
VP06, A11, and \citet{Denney12} demonstrate (1) that the scatter in the
\civ-to-\Hbeta\ masses or line widths is reduced when low-$S/N$ spectral
data are removed, and (2) that low $S/N$ can mask absorption in the
\civ\ line profile, leading to some of the highly discrepant
\civ-to-\Hbeta\ masses of N07 and \citet{Baskin&Laor05}.  Even without
these complicating issues, the uncertainty in the velocity field of the
BLR gas, as derived from a SE line-width characterization, is {\it
  already} the largest source of systematic uncertainty in SE mass
estimates due to the unknown geometry, kinematics, and inclination of
the BLR \citep{Woo10}.  When width measurements are routinely made from
survey data of varying quality, these uncertainties are enhanced.  These
fractional velocity errors are magnified in the mass estimates that
depend on $\Delta V^{2}$.

SE BH mass measurements are used to draw conclusions about black hole
demographics, growth rates, BLR gas kinematics, accretion and feedback
physics, and the evolution of all these properties across cosmic time
\citep{Vestergaard&Osmer09, Kelly10, Conroy13, Shankar13, Trump13}.  A
clear understanding of the effects that data quality and analysis
practices have on the accuracy and precision of these line width
measurements and BH mass estimates is therefore crucial.  In this work,
we attempt to reconcile the evidence and arguments on both sides of this
debate as to the reliability of BH mass estimates based on \civ\ in the
context of data quality.  We first select one of the studies that (1)
conclude that \civ\ is a poor virial mass estimator based on a large
scatter between \civ- and \Hbeta-based black hole masses (albeit over a
very limited dynamic range in luminosity), and (2) base their
conclusion solely on survey-quality (i.e., typically low $S/N$) data of
the \civ\ line.  For this, we select the work of N07, who present a
sample of 15 high-$z$ quasars with \civ\ masses based on spectra from
SDSS and \Hbeta\ masses determined using Gemini Near-Infrared
Spectrograph observations.  By obtaining new, high-$S/N$ spectra of a
portion of this sample and combining it with other high-quality data
from the literature, we investigate whether data quality can explain the
discrepancy between \civ- and \Hbeta-based BH mass estimates.  We define
high-quality spectra as those with $S/N \geq$10 pixel$^{-1}$ measured in
an emission-line-free region of the continuum ($\sim$1450\AA\ or
$\sim$1700\AA\ in the rest frame).

In Section \ref{S_data}, we present spectra of six high-redshift quasars
from the N07 sample observed with the first of the Multi-Object Double
Spectrographs \citep[MODS1;][]{Pogge10} on the Large Binocular Telescope
(LBT) and give details of the additional samples we select from the
literature and public archives to increase our total sample size to 47
AGNs.  In Section \ref{S_FitsWidths} we describe how we fit the \civ\
line profiles, and Section \ref{S_widthsMasses} describes our line
width, luminosity, and BH mass measurements.  We then compare the \civ\
and \Hbeta\ masses derived from our high-quality sample in Section
\ref{S_impactSNonMass}. In Section \ref{S_discussion} we discuss the
impact data quality has on (1) the presence of absorption in the \civ\
profile, (2) the \civ\ line width measurements, and (3) the SE \civ\ BH
masses in relation to \Hbeta\ masses.  A summary and concluding remarks
are given in Section \ref{S_conclusion}.

\section{Sample Selection and Observations}
\label{S_data}

\subsection{N07 Sample}

\begin{figure*}
\epsscale{1.2}
\plotone{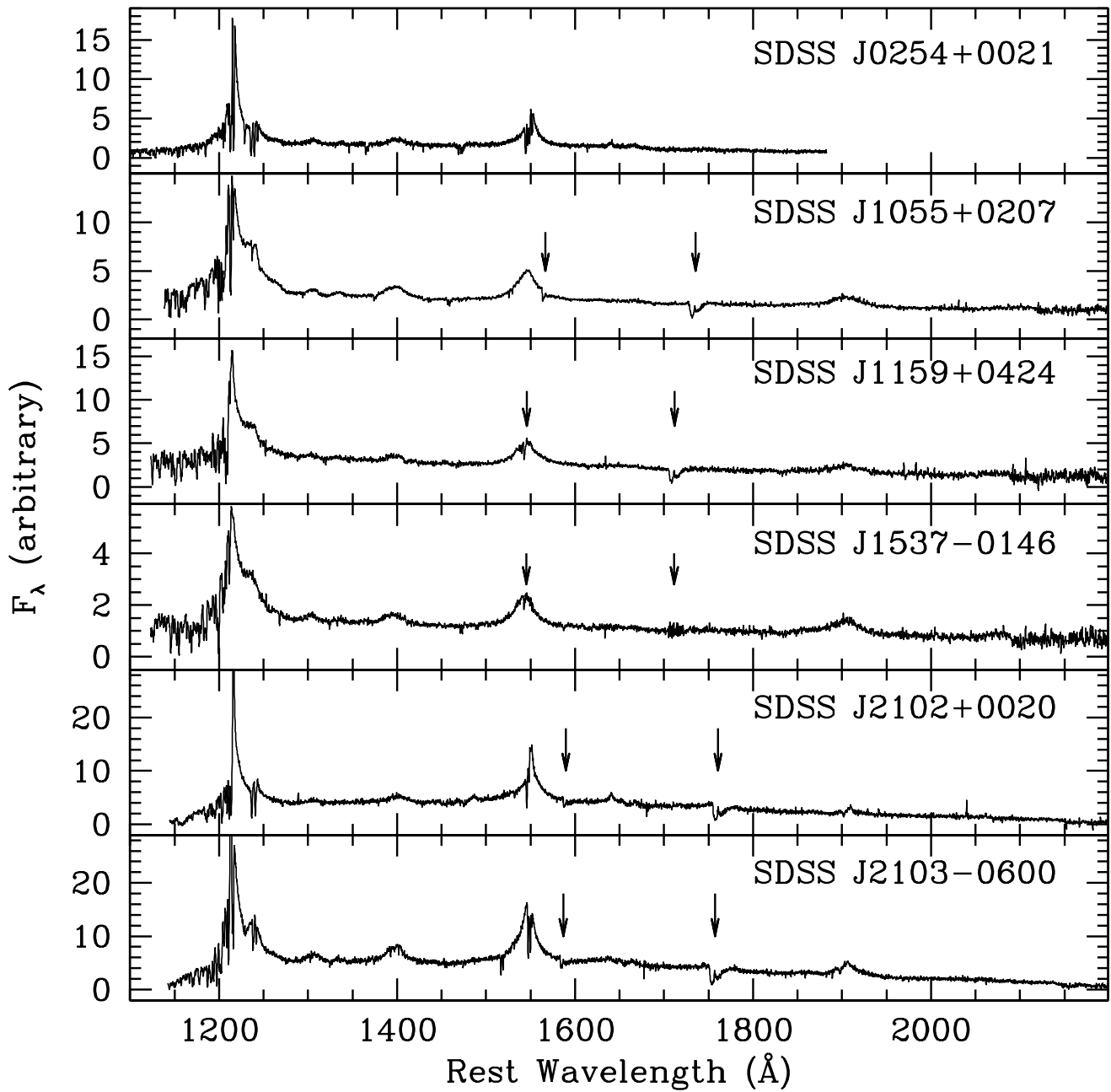}

\caption{Restframe UV spectra of the six high-redshift quasars observed
  with MODS1 on LBT.  These are all SDSS targets from N07.  The arrows
  indicate the location of atmospheric O$_2$ A-band and B-band
  absorption observed at observed wavelengths 7620\AA\ and 6880\AA,
  respectively.}

\label{fig:MODSfull}
\end{figure*}

We used the MODS1 spectrograph on the LBT to obtain rest-frame UV
spectra of six of the 15 high-redshift quasars presented by N07 (Table
\ref{Tab_Observations}; we will refer to targets by the leading digits
in their names, e.g., J0254 for SDSS J025438.37+002132.8). These six
quasars were chosen from the full N07 sample because of their favorable
location on the sky during our observing runs rather than their
particular spectral properties or \civ\ versus H$\beta$ mass estimates.

For each quasar we used either the red or blue channel of MODS1 without
the dichroic, depending on the observed wavelength of the redshifted
\ion{C}{4} $\lambda$1549 emission line. The blue-channel spectra used
the G400L grating (400\,lines\,mm$^{-1}$ in first order), and the
red-channel spectra used the G670L grating (250\,lines\,mm$^{-1}$ in
first order) with a GG495 order blocking filter. For all spectra we used
the 0\farcs6 segmented long-slit mask (LS5x60x0.6), centering the quasar
in the slit. This slit width gives a nominal resolution of
$\lambda/\Delta\lambda\approx$2000, with wavelength coverage from
3200$-$6000\AA\ in the blue channel and 6000$-$10000\AA\ in the red
channel. All of the quasars were observed near meridian crossing so we
did not need to orient the slit along the parallactic angle to minimize
the effects of differential atmospheric refraction (MODS does not have
an atmospheric dispersion corrector).  Multiple exposures (3 or 4) were
used to control for cosmic rays. The observations and observing
conditions are summarized in Table \ref{Tab_Observations}.

After processing the images using MODS-specific two-dimensional
calibration procedures (bias and flat field), the spectra were extracted
and then wavelength and flux calibrated using standard procedures in the
IRAF twodspec and onedspec packages. The spectral resolution of the
MODS1-Blue (-Red) channel targets is roughly 2.2\AA\ (3.5\AA) near the
\ion{C}{4} emission line. We resampled the spectra of J0254, J1055,
J1159, and J1537 onto a linear wavelengths scale with
0.5\AA\,pixel$^{-1}$, and J2102 and J2103 onto a linear wavelength scale
with 0.75\AA\,pixel$^{-1}$. Figure \ref{fig:MODSfull} shows the MODS1
spectra of our six targets.

Although spectrophotometric standard stars were observed as part of the
overall queue observing programs for the nights, the variable observing
conditions resulted in an unreliable absolute flux calibration.  Thus,
the spectra in Figure \ref{fig:MODSfull} are in uncalibrated
$F_{\lambda}$ units.  Unfortunately, telluric standards were also not
observed.  Telluric absorption is present to some degree in the \civ\
profiles of all targets except J0254; however, it is typically only
present in the line wing and does not hinder our ability to model the
line profile or measure the line width.  The exceptions to this are for
J1159 and J1537.  In the case of J1159, we did not include corrections
because of the proximity of the absorption to the profile peak plus the
additional intrinsic (or intervening) absorption.  For J1537, the
spectrophotometric standard taken the same night differed in airmass by
only $\sec(z)\sim0.1$.  We therefore performed a crude correction for
the O$_2$ A-band and B-band by dividing the standard star spectrum with
templates derived from the HST CALSPEC database and then using the NOAO
onedspec package {\small TELLURIC} task to scale the telluric
correction.  Some residual absorption at the optically thick, and
therefore nonlinear, core of the O$_2$ band head remains due to the
imperfect match in seeing and/or airmass, but we later masked any
remaining telluric absorption when we fit the \civ\ profiles (see
Section \ref{S_FitsWidths}).

In order to make a meaningful comparison with the original SDSS spectra,
we performed a homogeneous analysis of both the original SDSS spectra
and the MODS1 spectra of these six quasars.  The SDSS spectra were
rebinned to a linear wavelength dispersion consistent with the pixel
size at restframe $\lambda$1549\AA\ in each SDSS spectrum.  This
resulted in dispersions of 1.2\AA\ pixel$^{-1}$ for J0254, and 1.5\AA\
pixel$^{-1}$ for J1055, J1159, J1537, J2102, and J2103.  The spectral
resolution in this region was calculated to be 2.2--2.3\AA\ for all
sources.

\subsection{Additional Literature Samples}

\begin{figure*}
\epsscale{1.1}
\plotone{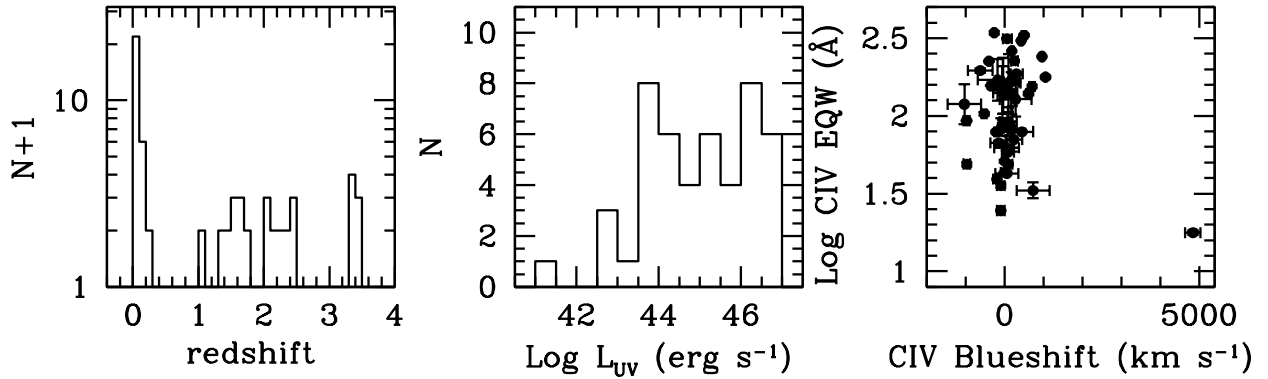}

\caption{Properties of our complete sample.  The left panel shows the
  distribution of redshifts, the middle panel shows the UV continuum
  luminosity distribution, and the right panel shows the location of our
  sample in the \civ\ blueshift--equivalent width parameter space
  described by \citet{Richards11}.}

\label{fig:samplestats}
\end{figure*}

We expand our sample of high $S/N$ spectra by including three additional
samples.  First, we use eight objects from A11, excluding the
broad-absorption line quasar H1413+117 and objects classified as having
Group II, poorer quality, line widths (see A11 Table 3 and Section 3).
Second, we include six of the 10 objects presented by D09 that fit our
quality requirement of $S/N>10$\, pix$^{-1}$ and do not have a broad
absorption-line region obscuring the blue side of the \civ\ profile.
Third, we fully reanalyze all the high $S/N$ UV spectra of the
reverberation mapping sample
\citep[e.g.,][]{Peterson04,Bentz08,Denney10,Grier12b} in the MAST
archives.  Much of this sample overlaps with that presented by VP06, but
we have updated it with recent high-resolution spectra taken with either
the {\it HST} Cosmic Origins Spectrograph (COS) or Space Telescope
Imaging Spectrograph (STIS).  In a limited number of cases, we averaged
multiple epochs that were closely spaced in time to increase the $S/N$,
but we otherwise dropped spectra/targets that did not meet our $S/N$
requirement, leaving 27 objects in the RM sample.  Our full sample
therefore contains 47 objects.  Figure \ref{fig:samplestats} shows the
redshift and UV luminosity distributions and the relation between \civ\
blueshift\footnote{Blueshifts were measured relative to the systemic
  redshift determined from the \ob\ emission lines.} and equivalent
width \citep[discussed by][]{Richards11} for our sample.  The former two
distributions (left and center panels of Figure \ref{fig:samplestats})
demonstrate that our sample, though not large by survey standards, spans
redshifts 0\ltsim\,$z$\ltsim\,3.5 and five orders of magnitude in AGN
luminosity.  The right panel shows that while this sample spans a broad
range of \civ\ equivalent widths (EQW), there is, unfortunately, still
only one object --- Q2302 from the D09 sample --- that has a low \civ\
EQW {\it and} a large blueshift.  It is, however, an extreme example of
this phenomenon.

\section{C\,{\small{IV}} Line Profile Fits}
\label{S_FitsWidths}

There is no universally accepted method for separating the various
blended components of emission in AGN spectra or for setting line
boundaries for measuring emission-line widths \citep[see, e.g.,
A11;][]{Vestergaard11, Park12a, Shen12}.  \citet{Denney09a} also
demonstrate that fitting functional forms to the \Hbeta\ profile can
exacerbate, rather than mitigate, systematic problems in the \Hbeta\
line width measurements.  In general, this is true only of low $S/N$
data, and is therefore not a concern here.  Regardless of data quality,
reliably measuring the \civ\ line widths {\it without} a model for the
intrinsic line shape is impossible in the presence of absorption in the
line profile and blending with the ``red shelf'' emission often seen
between \civ\ and \heii\ \citep[see][and references therein]{Fine10,
  Assef11}.  Using functional fits is therefore a common practice under
these circumstances, and thus, utilized here.

We chose a simple approach for fitting the \civ\ emission region that
closely follows the ``Prescription A'' approach described by A11 and the
continuum fitting of \citet{Fine10} methods (1) and (2).  We fit and
subtract a linear local continuum, fitting to a region blueward of \civ\
(rest wavelength $\sim$1450\AA, or $\sim$1350\AA\ in a few cases where
the $\sim$1450\AA\ region is contaminated by absorption) and redward of
\heii\, $\lambda$1640 and O\,{\sc iii}]\, $\lambda$1663 (at
$\sim$1700\AA).  By selecting the continuum windows this way, the red
shelf lies within our fitting window.  We do not assume an origin for
this emission for our fits.  Instead, we set our redward \civ\ line
boundary well into the red shelf and mask out the wavelength region
covered by the red shelf during the fit.  We select this region
independently for each object, but typically start the mask between
$\sim$1580$-$1600\AA\ and extend it to the red edge of O\,{\sc iii}],
$\sim$1690\AA. We also mask out any absorption regions and
\niv]\,$\lambda$1486 emission, when it is observed.

We then fit the unmasked regions of the continuum-subtracted \civ\ line
profiles with sixth-order Gauss-Hermite (GH) polynomials using the
normalization of \citet{vanderMarel93} and the functional forms of
\citet{Cappellari02}.  The best-fitting coefficients are determined with
a Levenberg-Marquardt least-squares fitting procedure.  Constraints from
the {\it unmasked} line and continuum regions provide interpolation
through the {\it masked} regions.  This approach minimizes the number of
components that are fit to the spectra, minimizing problems that can be
introduced by the use of multiple model fits to blended emission
components \citep[e.g.,][]{Denney09a}.

We do not constrain the number of GH fit components required to
reproduce the observed \civ\ profile, although typically, only two
components are required.  We do not attribute individual fit components
to kinematically distinct regions (e.g., narrow-line region (NLR) as
compared to BLR components).  Some studies \citep{Baskin&Laor05,
  Sulentic07, Greene10, Shen12} remove a narrow-line component, under
the assumption that this emission arises in a low-density, extended
kiloparsec-scale NLR.  Based on the arguments by \citet{Denney12},
including a demonstration that the non-reverberating, low-velocity
component of the \civ\ line is much broader than the \oiii $\lambda$5007
narrow emission line, we do not do so, and we use the full composite fit
for our \civ\ line width measurements.

\subsection{N07 Sample}

Using these procedures, we fit both the MODS1 and SDSS spectra of the
N07 sample.  Our goal in refitting the latter is (1) to attempt to
reproduce the line widths quoted by N07 based on these spectra, and (2)
to make a meaningful comparison between these survey quality spectra and
the high quality MODS1 spectra using the same modeling procedures.
While profile fits can introduce systematics into the measured \civ\
widths of the low $S/N$ SDSS spectra, direct measurements of either the
FWHM or line dispersion at these low signal-to-noise ratios are more
systematically uncertain than those made using parametric fits
\citep{Denney09a}.

\begin{figure*}
\figurenum{3a}
\epsscale{1.2}
\plotone{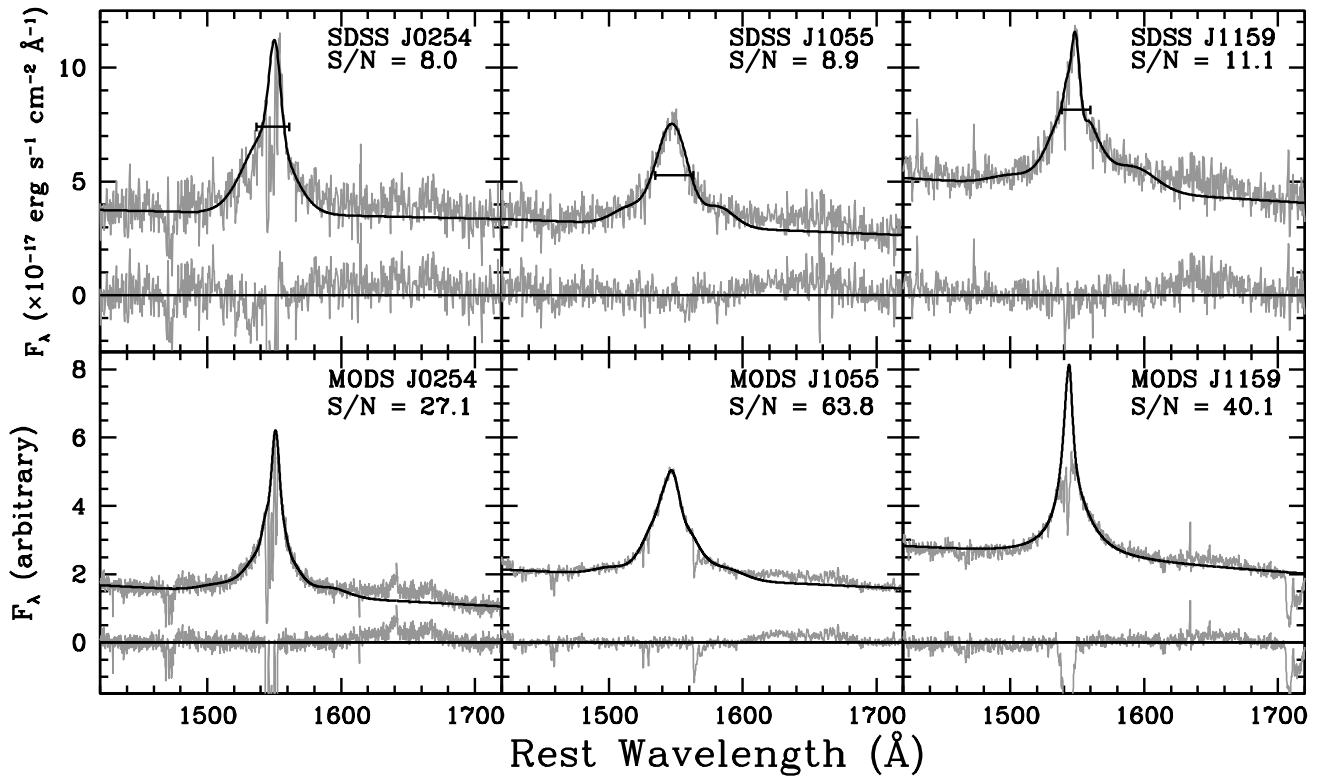}

\caption{Spectra (gray), model fits (solid black), and residuals for the
  N07 sample objects J0254, J1055, and J1159 observed with SDSS (top)
  and MODS1 (bottom).  The $S/N$ of each spectrum measured per
  resolution element in the continuum is given in the top corner of each
  panel.  The black bar across the SDSS \civ\ profile (top) shows the
  FWHM velocity width reported for the \civ\ emission line in this
  object by N07.}

\label{fig:MODSsampleCIVfits1}
\end{figure*}

Exceptions to this standard fitting procedure were required for the
MODS1 spectrum of J1159 and the SDSS spectrum of J0254.  The MODS1
spectrum of J1159 (Figure \ref{fig:MODSsampleCIVfits1}) shows
significant absorption across the peak (both intrinsic and atmospheric),
that prevents convergence to a physically-plausible GH-fit model.
Instead, we used multiple Lorentzian profiles to fit the wings of the
line profile, constraining the peak amplitude and wavelength by the
slope of the line wings and the functional form of the Lorentzian
profile.  For the SDSS spectrum of J0254, we also did not achieve an
acceptable GH polynomial fit, likely due to a combination of low $S/N$,
the apparent absorption, and asymmetry in this profile.  We obtained a
somewhat more representative profile with two Gaussian components, in
which the broader component is blueshifted relative to the narrower
component.

\begin{figure*}
\figurenum{3b}
\epsscale{1.2}
\plotone{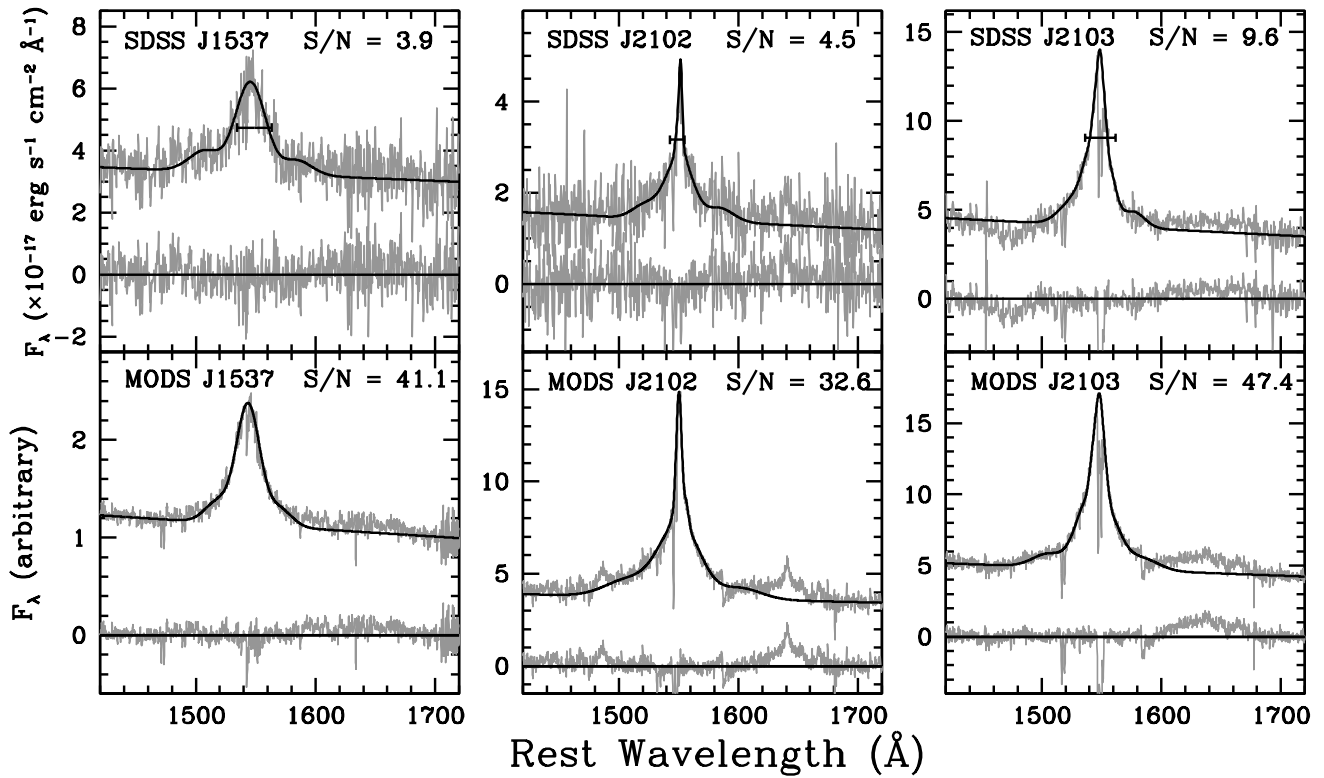}

\caption{Same as Figure \ref{fig:MODSsampleCIVfits1} but for the N07
  sample objects J1537, J2102, and J2103.}

\label{fig:MODSsampleCIVfits2}
\end{figure*}

The solid black curves in Figures \ref{fig:MODSsampleCIVfits1} and
\ref{fig:MODSsampleCIVfits2} show the final models (black) for each
spectrum (gray).  For comparison, the horizontal black bar above the
\civ\ profile in the top (SDSS) panels of Figures
\ref{fig:MODSsampleCIVfits1} and \ref{fig:MODSsampleCIVfits2} represent
the FWHM values given by N07.  These bars are centered at the half
maximum flux level and the theoretical \civ\ line center.  N07 do not
provide a description of how the line widths were measured or their
uncertainties, so a direct comparison is not possible.

\subsection{Additional Samples from the Literature}

\begin{figure*}
\figurenum{4}
\epsscale{1.2}
\plotone{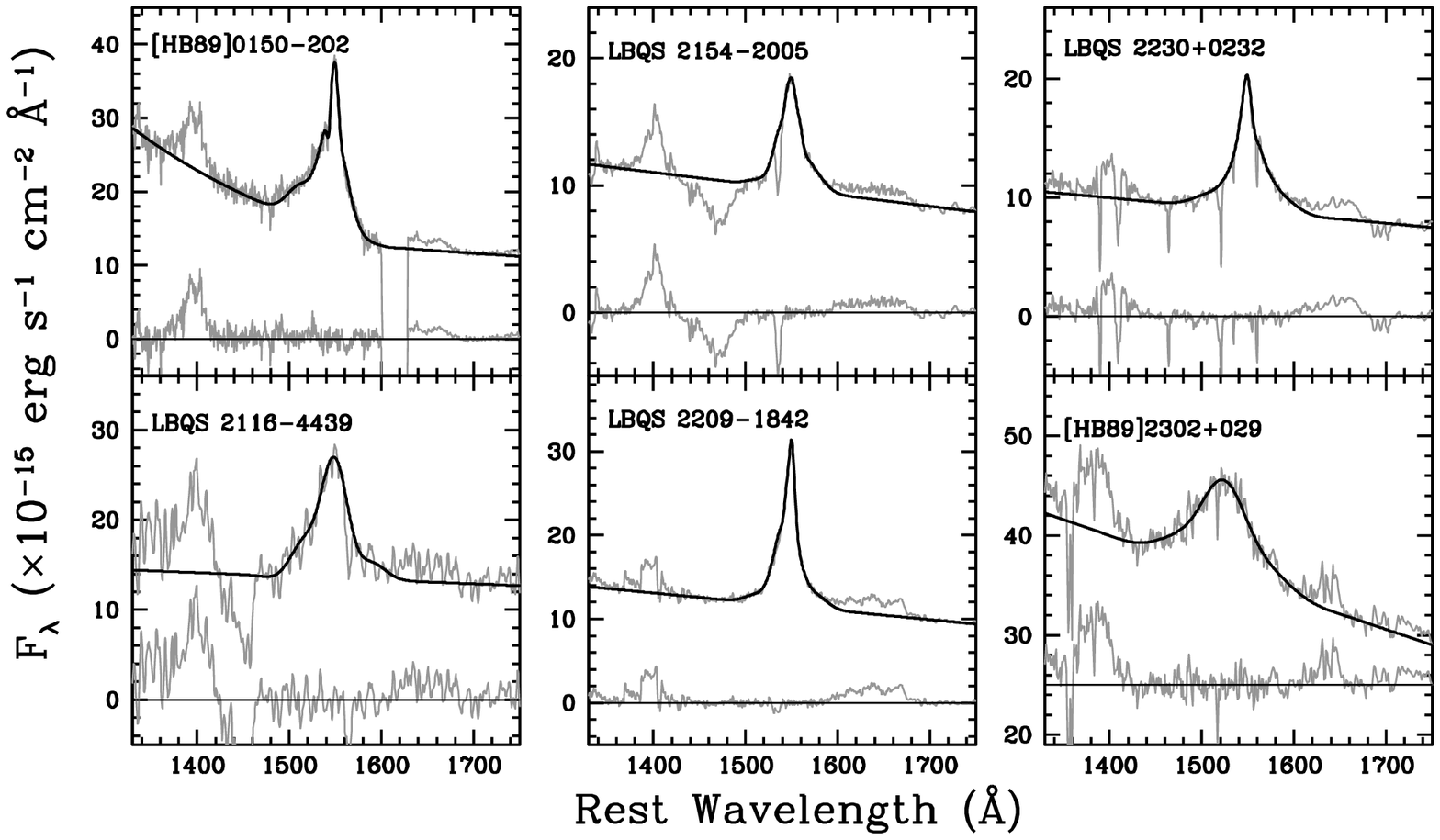}

\caption{Spectra (gray), composite continuum plus \civ\ profile fits
  (solid black), and residuals for the D09 sample.}

\label{fig:D09sample}
\end{figure*}

\begin{figure*}
\figurenum{5}
\epsscale{1.2}
\plotone{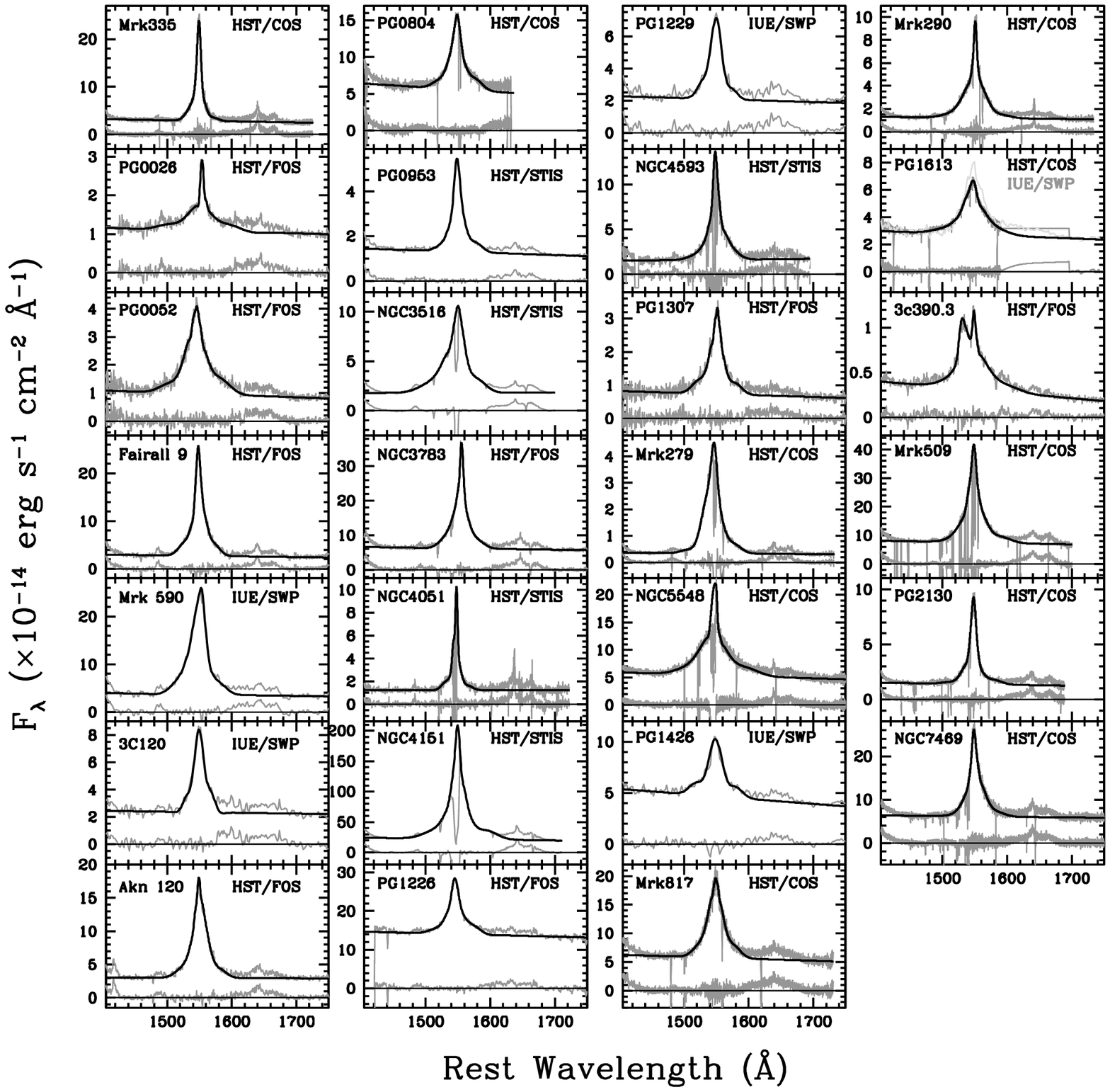}

\caption{Same as Figure \ref{fig:D09sample} but for the RM sample.}

\label{fig:RMsamplefits}
\end{figure*}

We fit the spectra for the other samples in the same standard manner
given above.  Our fits to the D09 and RM \civ\ spectra are shown in
Figures \ref{fig:D09sample} and \ref{fig:RMsamplefits}, respectively.
More than one high quality spectrum is available for many of the
objects in the RM sample, and in these cases we show a representative
example.  The \civ\ profile fits and exceptions to our standard fitting
method for the A11 sample can be found in that work.  The exceptions for
the D09 and RM samples are:

\noindent {\bf [HB89] 0150-202:} There is an unexplained, yet sharp,
difference in the continuum slope on either side of the dichroic, just
redward of the \civ\ line.  A linear continuum fit was therefore not
reliable, so we fit a local power-law continuum, based on regions
$\sim$1350\AA, $\sim$1450\AA, and $\sim$1700\AA.

\noindent {\bf PG0804:} The {\it HST}/COS spectrum did not cover the
$\sim$1700\AA\ continuum window.  We instead fit a linear continuum
between $\sim$1320\AA\ and $\sim$1450\AA\ and extrapolated it to the red
end of the available data.

\noindent {\bf PG1613:} The {\it HST}/COS spectrum did not cover the
$\sim$1700\AA\ continuum window.  However, there was also an {\it
  IUE}/SWP spectrum of this object.  We used the {\it IUE} spectrum as a
template for the $\sim$1700\AA\ continuum window of the COS spectrum by
scaling the {\it IUE} spectrum to match the $\sim$1450\AA\ continuum
flux of the COS spectrum.  We extrapolated the flux redward of the COS
spectrum using a constant value based on the last available pixel in the
original spectrum. This region was masked out during the \civ\ profile
fit.  We further extrapolated the COS spectrum to create a template
$\sim$1700\AA\ continuum window using the scaled {\it IUE} spectrum flux
values in this region.  Figure \ref{fig:RMsamplefits} shows both the
extrapolated COS spectrum (dark gray), scaled {\it IUE} spectrum (light
gray), and the resulting GH polynomial fit.  This extrapolation produces
a more symmetric and realistic \civ\ profile than extrapolating a linear
continuum fit between $\sim$1350\AA\ and $\sim$1450\AA, which resulted
in a much steeper slope redward of \civ\ than expected based on
comparison with the {\it IUE} spectrum.

\section{Line Width, Luminosity, and BH Mass Determinations}
\label{S_widthsMasses}

\subsection{H$\beta$}

\Hbeta\ line widths, optical luminosities or RM lag, and \Hbeta\ masses
were collected or recalculated from the literature as follows:

\noindent {\bf N07 sample:} We use the \Hbeta\ FWHM and 5100\AA\
monochromatic AGN luminosity given by N07 to recalculate the
\Hbeta-based masses directly from the calibration of the BLR $R-L$
relationship (\citealt{Bentz09rl}; Equation 4 of A11).  N07 does not
provide uncertainties in their measured quantities, so we have included
a `typical' uncertainty for the N07 \Hbeta\ masses of 0.2\,dex, based on
the typical uncertainties in the \Hbeta\ masses from the other samples
we consider.  Values are listed in Table \ref{Tab_N07Widths}.

\noindent {\bf A11 sample:} We adopt the \Hbeta\ masses for this sample
directly from Table 5 of A11, but we recalculate the uncertainties from
the line width and luminosity uncertainties given by A11 and the mass
scaling relation zero-point uncertainty of $\sigma_{\log \langle f
  \rangle}=0.09$ \citep{Woo10}.  A11 originally assigned uncertainties
to their masses to reflect the typically assumed global uncertainty in
SE mass estimates, which we now argue are too conservative (see Section
\ref{S_impactSNonMass}).  New uncertainties are on the order of
$0.1-0.3$\,dex.

\noindent {\bf D09 sample:} We use the \Hbeta\ FWHM and 5100\AA\
monochromatic AGN luminosity measurements from Tables 2 and 4 of D09 and
recalculate the \Hbeta\ mass using Equation 4 of A11.  Uncertainties are
calculated similarly to the A11 sample.  Values are listed in
\ref{Tab_D09widths}.

\noindent {\bf RM sample:} We use the direct RM-based \Hbeta\ mass
measurements for these objects, based on time delays measured from the
cross correlation method because these are the same measurements that
calibrated the $R-L$ relation \citep{Bentz09rl}, on which the other SE
\Hbeta\ masses are based \citep[see][for a complimentary method]{Zu11}.
For objects with multiple, reliable RM campaign measurements of the
\Hbeta\ time delay, we determine an error-weighted, geometric average of
the \Hbeta-based mass.  We also use the weighted uncertainty, which is
ultimately drawn from the measurement uncertainties in the RM lag, the
line width measured from the line dispersion of the \Hbeta\ profile in
the rms spectrum, and the uncertainty in $f$.  Table
\ref{Tab_Hbeta_masses} lists all of these values for this sample, and
the reader is referred to the original references listed there for
details of how the individual lag and line width measurements were made.

\subsection{C\,{\small{IV}}}

SE \civ\ masses are estimated from the mass scaling relationships in
Equations (7) and (8) of VP06, which require both a broad emission-line
width and a monochromatic UV continuum luminosity.  We characterize the
line width with both the FWHM and the line dispersion, $\sigma_l$, from
the continuum subtracted \civ\ profile fits (see Figures
\ref{fig:MODSsampleCIVfits1}--\ref{fig:RMsamplefits}) between the
spectral boundaries listed in Tables \ref{Tab_N07Widths},
\ref{Tab_D09widths}, and \ref{Tab_RMwidths}.  These widths were
corrected for spectral resolution following the procedures described by
\citet[][]{Peterson04} and references therein.  The line width
uncertainties were determined using the Monte Carlo approach of A11
based on 1000 resampled spectral models.  We describe the calculation of
the \civ\ masses for each sample below. Uncertainties were determined
using measurement uncertainties on the line widths and luminosities and
on the fit uncertainty in the mass scaling relation zero-point:

\noindent {\bf N07 sample:} For the SDSS spectra, we use the remeasured
line widths from this work combined with the 1350\AA\ monochromatic
luminosities given by N07.  For the MODS1 spectra, we use the line
widths measured here and the same luminosity as that used for the
SDSS-based masses.  We accounted for the possibility of intrinsic
variability by adding uncertainties to the luminosities of 0.08 dex.
This is based on the expected level of variability of approximately 0.2
mag estimated from the structure function of SDSS quasars
\citep{Vandenberk04, Macleod10}, over the time separating the
observation dates for the SDSS and MODS1 spectra (typically 2--3 years
in the quasar rest frame).  Line widths, luminosities, and other
properties of this data set are listed in Table \ref{Tab_N07Widths}.

\noindent {\bf A11 sample:} The line widths and masses were calculated
by A11 in a manner consistent with our present methods.  We therefore
use the exact line widths and masses presented in that work, and readers
are referred to Tables 3 and 5 of A11 for all measurements related to
this sample.  However, the \civ\ mass uncertainties we adopt for the A11
sample have been recalculated as described above.

\noindent {\bf D09 sample:} We use the 1350\AA\ monochromatic
luminosities given by D09 and the new \civ\ line widths measures from
our fits to these data.  See Table \ref{Tab_D09widths} for these
measurements.

\noindent {\bf RM sample:} Line widths are measured from the \civ\ fits
to all epochs for all objects, 64 in total.  Monochromatic luminosities
are also measured from the same data after correcting for Galactic
extinction \citep{Schlafly11}, where we adopt the luminosity measured at
1450\AA\ because this is typically a cleaner continuum window in our
data, and VP06 demonstrate it to be equivalent to that at 1350\AA.
Uncertainties in the luminosity were determined from the standard
deviation of the luminosities measured from the resampled spectral
models used for estimating uncertainties in the line widths.  For
objects with multiple epochs, we calculate the FWHM- and
$\sigma_l$-based \civ\ mass from the uncertainty weighted geometric mean
from all SE (or RM campaign mean spectrum) mass estimates.  The
uncertainty in this mean mass is taken to be the quadrature sum of the
standard deviation about the unweighted mean mass and the weighted
measurement uncertainty on the weighted mean mass.  This takes into
account intrinsic variability effects between the multiple SE \civ\
measurements to which the direct RM-based \Hbeta\ mass is not
susceptible.  Adopting error bars that account for both measurement
uncertainties and intrinsic variability more accurately reflect the
limitations in the precision with which we can measure a SE BH mass for
this relatively lower-luminosity sample, where short time-scale
variability could introduce a source of scatter not likely to be of
significance for quasars.  Line widths, luminosities, \civ-based masses
and uncertainties and other spectral properties are listed in Table
\ref{Tab_RMwidths} for the RM sample.

\section{Comparison between C\,{\small{IV}} and H$\beta$ BH
  Masses}
\label{S_impactSNonMass}

\begin{figure}
\figurenum{6}
\epsscale{1.2}
\plotone{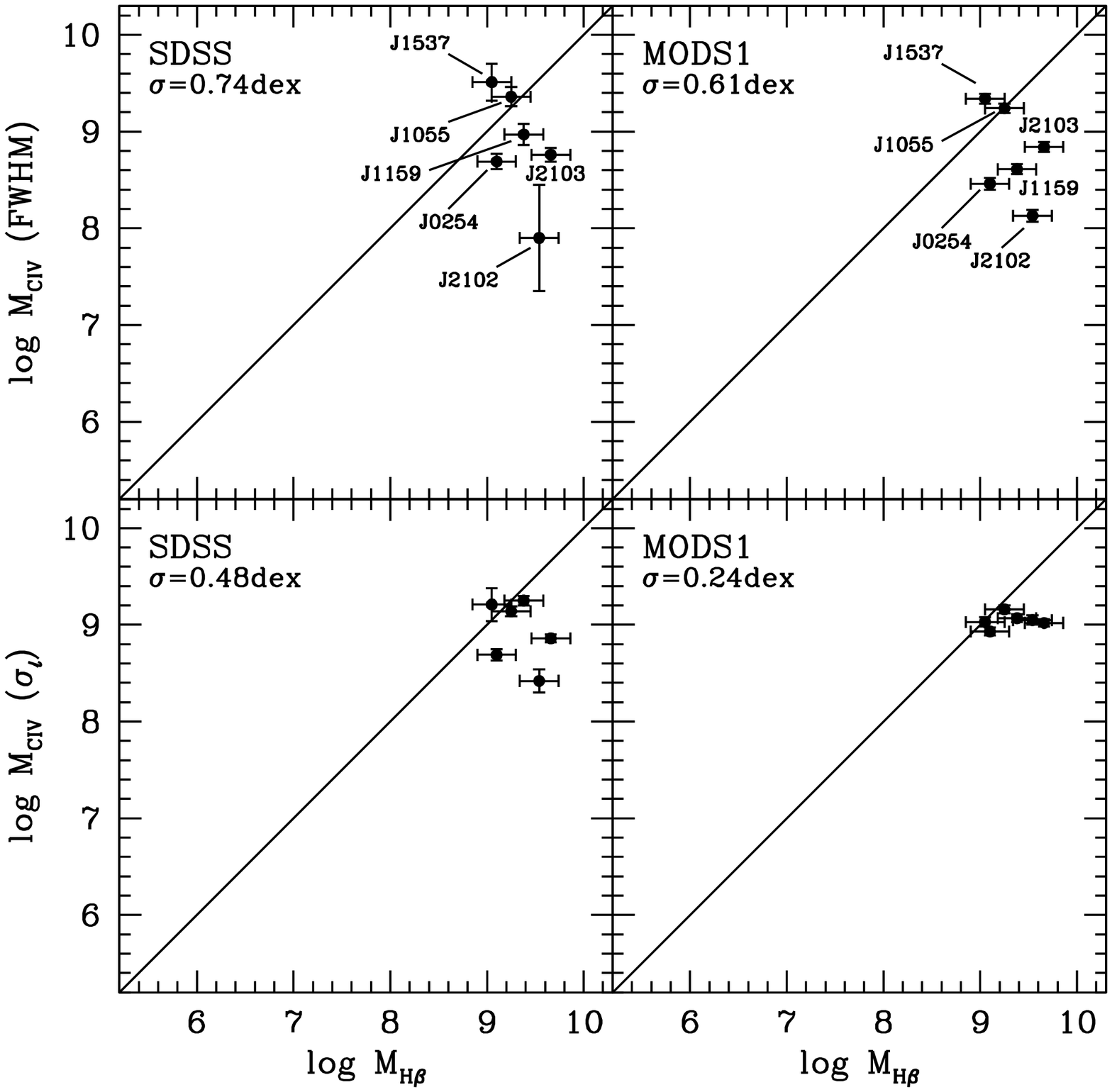}

\caption{Comparison between \Hbeta\ masses and \civ\ masses estimated
  from the SDSS spectra (left) and the MODS1 spectra (right) of the N07
  sample.  Top panels show \civ\ masses determined with the FWHM and
  bottom panels show masses based on $\sigma_l$.  The solid line shows
  where the \civ\ and \Hbeta\ masses are equal.  The scatter, quantified
  as the standard deviation about the mean of the \civ-to-\Hbeta\ mass
  residuals, $\sigma$, is shown in the top left of each panel.
  Individual objects are labeled in the top panels to simplify a
  comparison between the \civ\ masses of each data set.}

\label{fig:N07samplemasscompare}
\end{figure}

We can now compare the \civ\ SE masses with the \Hbeta\ masses.  We
focus first on the N07 sample, for which we have both low and high
quality spectra of the same objects.  These are high-luminosity,
high-redshift sources, so intrinsic variability is unlikely to
significantly impact a comparison between the \civ\ and \Hbeta\ masses.
Figure \ref{fig:N07samplemasscompare} compares \civ\ masses calculated
from the SDSS spectra (left panels) and the MODS1 spectra (right panels)
determined with the FWHM (top panels) and $\sigma_l$ (bottom panels) to
the \Hbeta\ masses.  A distinct correlation between the \civ\ and
\Hbeta\ masses is not apparent for this small sample, but this is not
surprising given the small dynamic range in mass.  Despite this small
sample, we can still postulate two consequences of the differing data
quality between the SDSS and MODS1 spectra:

\begin{enumerate}

\item Data quality does {\it not} improve, at least significantly, the
  consistency between \civ\ and \Hbeta\ masses when using the FWHM to
  characterize the \civ\ line width.  Instead, an equally poor
  correlation with a large scatter, $>$0.6\,dex, is found between both
  low-quality SDSS spectra and high-quality MODS1 spectra.

\item Data quality {\it does} improve the consistency of \civ\ and
  \Hbeta\ masses when using $\sigma_l$ to characterize the \civ\ line
  width.  The improved $S/N$ of the MODS1 spectra over that of the SDSS
  spectra reduces the scatter in the mass residuals, $\log M(\civm)-\log
  M({\rm H}\beta)$, by a factor of 2, from 0.48\,dex to 0.24\,dex.

\end{enumerate}

\begin{figure}
\figurenum{7}
\epsscale{1.0}
\plotone{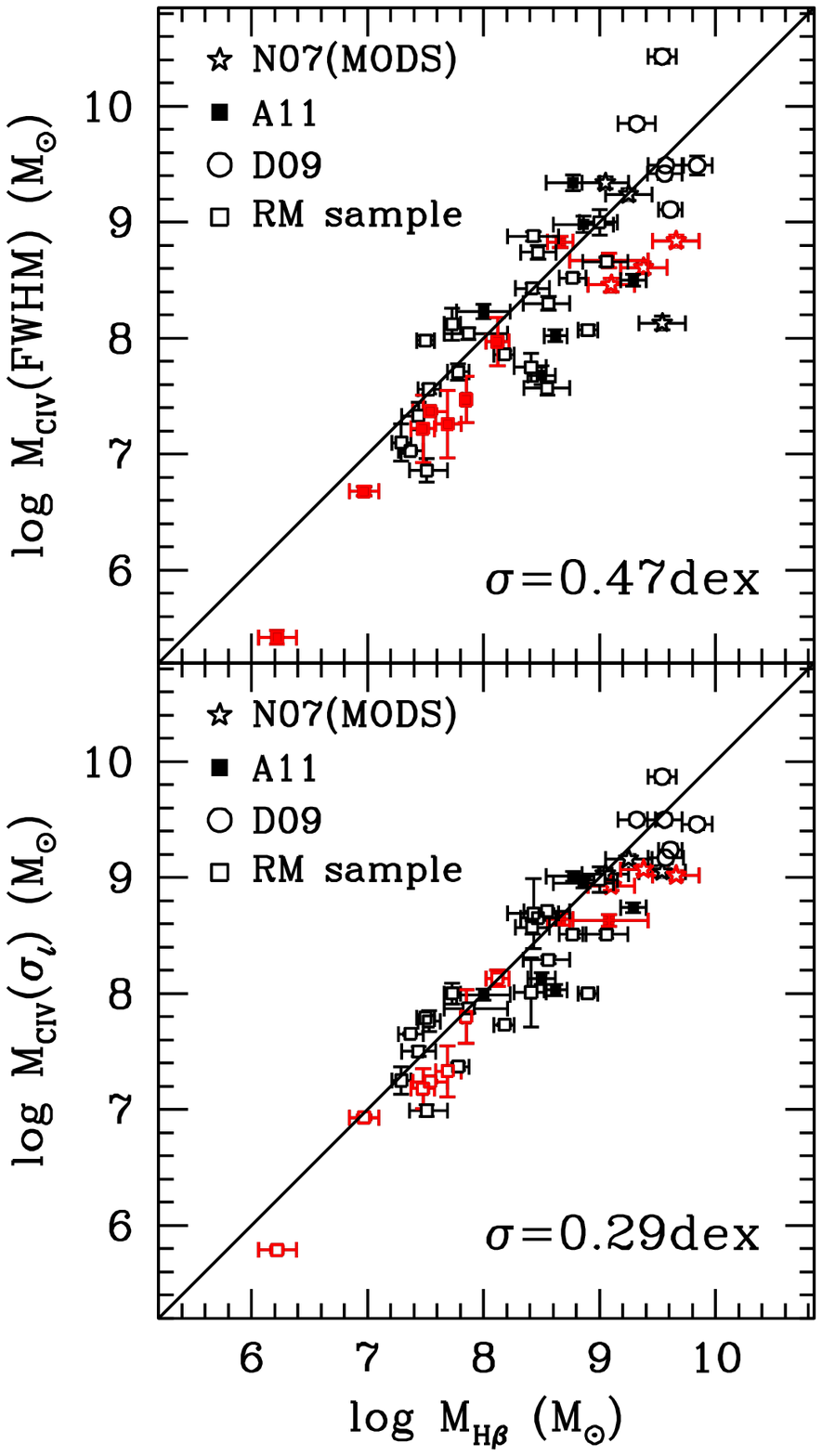}

\caption{Comparison between \Hbeta\ masses and \civ\ masses estimated
  from our complete sample of high quality data.  The top (bottom) panel
  shows \civ\ masses based on the FWHM ($\sigma_l$).  The solid line
  shows where the \civ\ and \Hbeta\ masses are equal.  The scatter,
  $\sigma$, quantified as the standard deviation about the mean of the
  sample of \civ-to-\Hbeta\ mass residuals, $\log M(\civm)-\log M({\rm
    H}\beta)$, is shown in the bottom right of each panel.  The 12 red
  points represent objects for which absorption was observed across the
  peak of the \civ\ emission line.}

\label{fig:masscomparefullsample}
\end{figure}

Although this direct comparison of data quality effects is useful, this
sample is too small and limited in dynamic range for deriving general
conclusions.  Figure \ref{fig:masscomparefullsample} shows the \civ\
masses based on the FWHM (top panel) and $\sigma_l$ (bottom panel)
against the \Hbeta\ masses for our full sample of 47 high quality
spectra from the N07, A11, D09, and RM samples.  The effects implied
from Figure \ref{fig:N07samplemasscompare} are now clearly apparent.
This larger sample now spans 4\,dex in BH mass, so there is a clear
correlation between \civ\ and \Hbeta\ mass.  However, even with
high-quality data, there is significant scatter between the FWHM-based
\civ\ mass measurements and their \Hbeta\ counterparts.  The standard
deviation about the mean of the sample of FWHM-based \civ-to-\Hbeta\
mass residuals is $\sigma=0.47$\,dex.  On the other hand, the scatter
observed between \Hbeta\ masses and \civ\ masses derived from the line
dispersion, $\sigma_l$, in high quality data is only
0.29\,dex\footnote{The scatter in both panels of Figure
  \ref{fig:masscomparefullsample} was calculated excluding 3C 390.3, the
  largest outlier in the bottom panel.  Two independent \Hbeta\ RM
  results exist for this object \citep{Dietrich98, Dietrich12}.  The
  measured time delays ($R_{\rm BLR}$) and luminosities between the two
  campaigns behaved as expected from photoionization physics ($R\sim
  L^{-1/2}$).  However, the \Hbeta\ velocity widths defied the virial
  expectations ($\Delta V\sim R^{-1/2}$, and thus $\Delta V\sim
  L^{-1/4}$) --- the measured line widths were larger when the object
  was in a higher luminosity state.  Consequently, the RM BH mass
  measurement differed by approximately an order of magnitude between
  the two campaigns.  Furthermore, this object exhibits complex, often
  double- or even triple-peaked broad emission line profiles, raising
  questions as to the best way to define a characteristic, mean velocity
  from such complex profiles.  For this reason, the \Hbeta\ mass, and
  therefore possibly the \civ\ mass, for this object is likely
  unreliable.}.  There is also a zero-point offset between the observed
mean of each \civ\ mass distribution and that of equality with the
\Hbeta\ masses.  This offset is related to the zero-point calibration of
the SE \civ\ mass scale taken from VP06 and is simply due to the
prescriptional differences between our line width measurements and those
of VP06.  This type of zero-point calibration issue does not affect our
results.  In addition, while we have taken care to place all of the
\Hbeta\ masses on the same mass scale, the line widths and luminosities
or lags were taken from the literature and were not measured with a
homogeneous method.  This likely adds additional scatter to the
comparisons shown in Figure \ref{fig:masscomparefullsample} that is not
associated in any way with \civ.  Nonetheless, since both the top and
bottom panels use the same \Hbeta\ masses, this does not affect the
relative difference in scatter between the \civ\ FWHM- and
$\sigma_l$-based masses shown here.

\section{Discussion}
\label{S_discussion}

A simple qualitative comparison of the noisy SDSS and MODS1 spectra
shown in Figures \ref{fig:MODSsampleCIVfits1} and
\ref{fig:MODSsampleCIVfits2} clearly demonstrates the deleterious
effects low quality data can have on our ability to accurately describe
quasar emission line properties.  Using this comparison and our results
above, we discuss the effects of data quality on characterizing
absorption, the \civ\ velocity width, and SE \civ\ BH mass estimates.

\subsection{Absorption}

One striking characteristic of even the small sample of objects observed
with both SDSS and MODS1 is the prevalence of narrow absorption features
in the \civ\ profiles.  Because \CIV\ is a resonance transition,
self-absorption is common ($>$50\%), and it is usually in the form of
narrow absorption lines (NALs) due to gas associated with the quasar
and/or along the line of sight \citep[e.g.,][]{Vestergaard03, Wild08,
  Gibson09, Hamann11}.  Recognizing the presence and extent of NAL
features in low quality data with a high level of confidence is
difficult or impossible.  This was demonstrated by A11 for another of
the N07 targets (SDSS1151+0340) where an absorption feature was missed
in the SDSS spectrum.  We see here that the opposite is also possible,
as we misidentified noise in the SDSS spectrum of J0254 (Figure
\ref{fig:MODSsampleCIVfits1}, top left) near 1510\AA\ and 1515\AA\ as
absorption, and our fit was affected by this assumption.  Correctly
identifying and modeling intrinsic absorption is absolutely necessary
for measuring accurate \civ\ BH masses.

With high $S/N$ and relatively high spectral resolution data, where the
{\it presence} of absorption can be accurately detected, the absorption
features can usually be masked and interpolated across with relatively
few consequences for the line widths and masses.  However, when the
absorption occurs at the peak of the \civ\ emission line, it is
difficult to know how well an arbitrarily defined profile based on a
functional fit reproduces the intrinsic emission-line profile and peak
amplitude.  There is simply no a priori expectation for the detailed
line shapes of individual AGNs.  We have marked objects with absorption
observed across the \civ\ emission line peak as red points in Figure
\ref{fig:masscomparefullsample} to show the possible contribution of
line peak absorption to the observed scatter in the masses.  We find
that the distribution of objects with absorbed peaks is not
systematically different that of the unabsorbed objects with respect to
the mean \civ\ to \Hbeta\ mass ratio, and the scatter in the \civ\ to
\Hbeta\ mass residuals found for our sample actually marginally
increases upon omission of these objects from the full sample, from 0.47
to 0.49\, dex and 0.29 to 0.30\, dex for FWHM- and $\sigma_l$-based
\civ\ masses, respectively.

The change in scatter after omitting the 12 absorbed-peak objects is
small and could simply be due to small number statistics.  In general,
however, masses estimated from absorbed-peak profiles using $\sigma_l$
will be less prone to biases in the width measurement than FWHM-based
mass estimates, because of the relative insensitivity of the line
dispersion measurement to the profile peak amplitude.  In contrast,
absorption and the resulting interpolation uncertainties across the
profile peak is more likely to bias FWHM measurements, which {\it are}
very sensitive to the amplitude of the emission-line peak.  A likely
explanation for the lack of additional scatter (and even a slight
reduction in the scatter) in the masses because of absorbed-peak
objects, here, is that the \civ\ line peak is already contaminated by
the non-variable emission component described by \citet{Denney12}.
Interpolating over the absorbed peak with Gaussian or Gauss-Hermite
functions is more likely to underestimate than overestimate the peak
amplitude \citep{Denney09a}, particularly for the relatively more
contaminated, `peaky' \civ\ profiles.  This would, fortuitously, reduce
the contamination of this component to the width measurement, leading,
in these random cases, to a more accurate \civ\ mass estimate.  This
should be the case, in general, but is even more likely to occur with
FWHM-based masses, and we do measure the marginally larger difference in
scatter in the FWHM-based masses between inclusion or omission of the
absorbed-peak objects.

\subsection{Line Profile and Width Characterization}

\begin{figure*}
\figurenum{8}
\epsscale{1.0}
\plotone{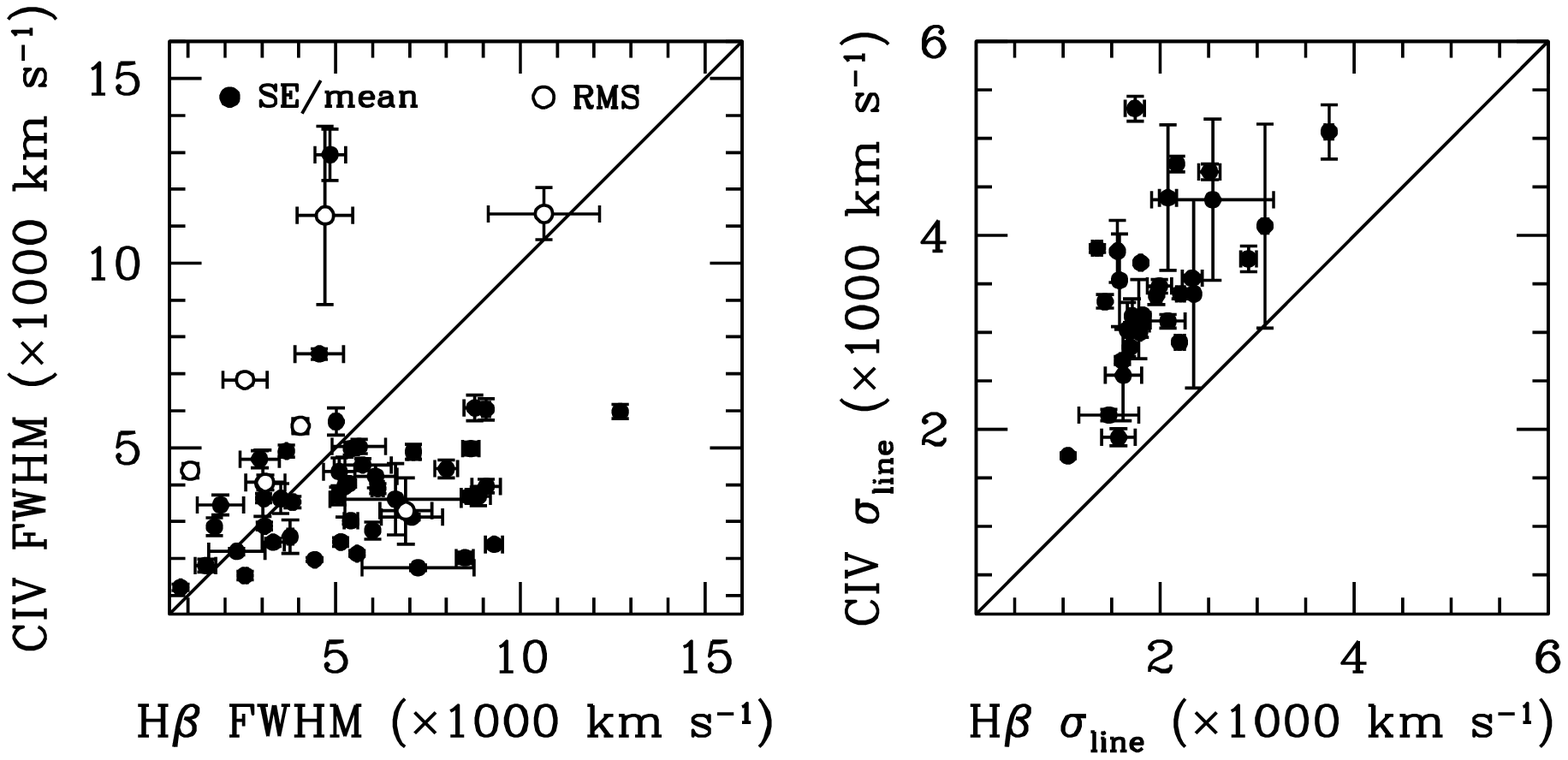}

\caption{{\it Left}: Comparison between \civ\ and \Hbeta\ FWHM
  measurements.  {\it Right}: Comparison between \civ\ and \Hbeta\
  $\sigma_l$ measurements for objects where \Hbeta\ $\sigma_l$ was
  available.}

\label{fig:widthcompare}
\end{figure*}

The reliability of \civ\ masses has been debated in the literature not
only as a result of the large scatter typically observed between \civ\
and \Hbeta\ masses themselves, but also because of the lack of
correlation observed between \civ\ line widths and those of other broad
emission lines \citep[see, e.g.,][]{Baskin&Laor05, Ho12, Shen12,
  Benny&Netzer12}.  The closed points in the left panel of Figure
\ref{fig:widthcompare} shows a similar result from our own data.  Here
we have used the \Hbeta\ line widths from \citet{Collin06} for the mean
spectrum of the RM sample objects in order to be more consistent with
using a SE \civ\ line width.  Also, we adopt the mean and standard
deviation of the \civ\ FWHM for objects with more than one epoch of
data.  In general, the \civ\ line width is expected to be broader than
\Hbeta\ because of the ionization stratification of the BLR.  However,
the values shown in the left panel of Figure \ref{fig:widthcompare}
scatter both above and below the line of equal \civ\ and \Hbeta\ widths.
This indicates that characterizing the high-ionization BLR gas velocity
using the \civ\ FWHM from a SE spectrum does not support the physical,
virial expectations from an ionization-stratified BLR under the
gravitational influence of the BH.  On the other hand, the open points
compare the \civ\ and \Hbeta\ FWHM measured in the rms spectrum and
taken from \citet{Denney12} and \citet{Collin06} for each line,
respectively, for the six objects that have both \civ\ and \Hbeta\ RM
measurements: 3c390.3, Fairall 9, NGC\,3783, NGC\,4151, NGC\, 7469, and
NGC\,5548 (for which we have two independent measurements of both \civ\
and \Hbeta).  When sampling only the reverberating gas velocities, the
expectation from a virialized and ionization stratified BLR holds for
all objects but Fairall 9, but in this case, the rms \civ\ profile is
weak and likely contaminated by noise, so the FWHM measurement should
not be trusted \citep[see][]{Rodriguezpascual97, Denney12}.

Similar studies in the literature have exclusively focused on comparing
FWHM measurements.  However, the right panel of Figure
\ref{fig:widthcompare} shows a similar width comparison using the line
dispersion to characterize both the \Hbeta\ and \civ\ line widths for
the objects with available \Hbeta\ $\sigma_l$ measurements.  In this
case, the relation between \civ\ and \Hbeta\ velocities {\it do} follow
the virial expectation, even with using the SE line widths: the \civ\
line widths are exclusively larger than the \Hbeta\
widths\footnote{Characterizing the BLR velocity field using $\sigma_l$
  also shows the greatest consistency with virial expectations for the
  correlation between the BLR velocity and radius \citep{Peterson04}.
  It is thus the preferred line width characterization for RM studies.}.
Note, however, that a tight correlation is not necessarily expected here
because the \civ\ and \Hbeta\ widths are not measured from simultaneous
epochs; intrinsic variability creates scatter, since virial expectations
imply $\Delta V \propto L^{-1/4}$, and $\Delta L(t)$ is significant
($>$1\,dex) for some of the RM sample objects between the \civ\ and
\Hbeta\ observations.  The level of scatter in Figure
\ref{fig:widthcompare} is therefore likely inflated, compared to typical
expectations for high-$L$ QSOs, because the sample is dominated by the
lower-luminosity, more variable RM sample.  An additional consideration
is that the ratio between the \civ\ and \Hbeta\ velocity widths may
depend on the specifics of the BLR structure and accretion rate, leading
to some intrinsic scatter between objects.  RM time delays for the few
objects with both \Hbeta\ and \civ\ results show that the \civ\ response
is typically $2-3$ times shorter than the \Hbeta\ response, and the lag
and line width follow virial expectations \citep{Peterson00a,
  Peterson04}, but this sample consists of only a handful of
intermediate luminosity AGNs.

An additional argument against using the FWHM to derive SE \civ\ masses
was presented by \citet{Denney12} and follows from the results in the
left panel of Figure \ref{fig:widthcompare}.  \citet{Denney12}
demonstrated that there is an emission component in the SE \civ\ profile
that is non-variable and therefore does not seem to be emitted from the
same velocity distribution of BLR gas that reverberates in response to
the continuum emission.  This non-variable component is likely
responsible for the `peaky' low-velocity core seen in many \civ\
profiles (although the non-variable emission may also be present in an
additional, or alternate, blue-shifted, broader component).  Its
presence can significantly contaminate the BLR velocity width
measurement when characterizing the SE \civ\ line profile with the FWHM.
This bias is likely responsible for much of the excess scatter seen in
our sample when using the FWHM to derive \civ\ masses and explains why
the \civ\ FWHM measured from the SE spectrum is often narrower than the
\Hbeta\ FWHM (although Figure \ref{fig:widthcompare} demonstrates the
same is not true when using the rms spectrum where this component is not
present). Thus, despite the problem of absorption, it appears that data
quality is not the {\it leading} cause of scatter between FWHM-based BH
mass measurements, though there is some effect (see Figure
\ref{fig:N07samplemasscompare} and \citealt{Denney09a}).
 
The line dispersion, $\sigma_l$, is not as sensitive to the line peak as
the FWHM and is therefore less affected by any contamination from the
non-variable \civ\ emission component (although strictly speaking, there
must be an effect on some level).  The additional insensitivity of
$\sigma_l$ to absorption in the line peak is another advantage of this
line width characterization.  However, $\sigma_l$ is sensitive to
correctly characterizing the wings of the lines, and as such, is very
sensitive to $S/N$.  In low $S/N$ spectra, it is difficult, if not
impossible, to accurately define the boundaries of the emission line and
characterize the intrinsic line shape as it merges with a noisy
continuum level.  A comparison between the SDSS and MODS1 spectrum of
J2102 shown in Figure \ref{fig:MODSsampleCIVfits2} clearly demonstrates
the improved clarity with which the wings and extent of the \civ\
profile can be distinguished from the continuum in high $S/N$ data as
opposed to survey-quality data.  Unfortunately, the limiting $S/N$
required to accurately trace the intrinsic profile is somewhat dependent
on the \civ\ line shape.  The more `boxy', low equivalent width
profiles, such as J1055 and J1537, can be more accurately fit in lower
$S/N$ data than the `peaky', extended-wing profiles like J2102 because
of the relative extension and contrast of the wings compared to the
noise level in the continuum.

Mass estimates based on $\sigma_l$ are clearly superior to those based
on the FWHM with the caveat that $\sigma_l$ is sensitive to blending in
the line wings, so high quality data are required.  This can be a
significant source of bias in using $\sigma_l$ for \Hbeta\ widths
\citep{Denney09a}.  This could be a source of bias for \civ\ as well, as
the source of the blended red shelf emission is still uncertain, and
misattributing the origin of this emission could bias the resulting
\civ\ $\sigma_l$ measurement \citep[see][]{Fine10,Assef11}.
Nonetheless, applying a homogeneous spectral decomposition and line
width measurement procedure to measure $\sigma_l$ can produce consistent
$\sigma_l$ measurements that lead to little scatter in \civ\ mass
estimates as compared to \Hbeta. As usual, care must be taken in
combining samples in order to mitigate scatter resulting from the use of
different methodology.

\subsection{Black Hole Masses}
\label{S_BHmasses}

Higher data quality makes a clear positive impact on the consistency
between SE \civ\ and \Hbeta\ BH masses when using the line dispersion to
characterize the \civ\ velocity field.  This is demonstrated by the
significant reduction in scatter between the low (SDSS) and high (MODS1)
quality $\sigma_l$-based \civ\ masses of the N07 sample (see Section
\ref{S_impactSNonMass} and Figure \ref{fig:N07samplemasscompare}).
Furthermore, the scatter of only 0.29\,dex between the \civ\
$\sigma_l$-based masses and \Hbeta\ masses, measured across our full
sample of {\it high quality} data shown in Figure
\ref{fig:masscomparefullsample}, is arguably the lowest so far quoted in
the literature between \civ\ and \Hbeta\ masses for a sample this size,
particularly since this (1) does not depend on any type of empirical,
potentially sample-dependent, correction, (2) does not factor in the
evidence from A11 that a continuum color correction is a comparable
contributor to this scatter, and (3) does not take into account
inhomogeneities in how the \Hbeta\ line widths were measured or other
systematics that may be associated with the \Hbeta\ mass estimates,
which is outside the scope of this work.  Another relatively direct
piece of evidence for the ability of data quality to reduce the scatter
between $\sigma_l$-based \civ\ masses is to look at the RM sample.  We
measure this scatter to be 0.28\,dex in our high-quality RM sample.
VP06, whose sample largely overlaps with our own\footnote{We use 24 of
  the 27 targets presented by VP06.  Mrk\,79, Mrk\,110, and PG1617+175
  were dropped from our analysis due to the unavailability of
  high-quality UV data.  However, we additionally include PG0804+761,
  NGC\,4593, and Mrk\,290, for which new high-quality UV data and/or RM
  results have become available.  Our RM sample is thus the same size as
  that of VP06.} but is more heterogeneous in quality, quote a scatter
of 0.37\,dex when using weighted averages of the multiple SE \civ\
masses and \Hbeta\ RM campaign masses.  The difference between our
results and that of VP06 is due in part to data quality differences
\citep[see also discussion by][]{Denney12}.  There are also differences
in our spectral analysis and line width measurement methods compared to
those of VP06 that may contribute to the reduced scatter.  We have
learned a lot about the sources of systematic problems in line width
measurements since the VP06 study \citep[e.g.,][]{Denney09a, Fine10}, so
it is not surprising that our reanalysis of this sample has fewer
systematic problems.

Conversely, when considering FWHM-based SE \civ\ mass estimates, our
results indicate that obtaining high quality data only marginally
improves the consistency between \civ\ and \Hbeta\ SE mass estimates.
Figure \ref{fig:N07samplemasscompare} shows a consistently large scatter
between \Hbeta\ and FWHM-based \civ\ masses for both low and high
quality data, and our results from the top panel of Figure
\ref{fig:masscomparefullsample} including the full sample of
high-quality data corroborate this finding.  We can again make a
comparison to the VP06 results for the RM sample to look at the
differences in scatter between our high-quality data set and their
heterogeneous-quality data set.  VP06 quote a scatter of 0.43\,dex,
while we find a scatter of 0.36\,dex.  There is some improvement, but
again prescriptional differences could play a part in this difference as
well.

Finally, it is still difficult with our sample to address the concerns
of \citet{Richards11} regarding the applicability of existing \civ\ SE
mass scaling relationships to quasars covering the complete \civ\
EQW--blueshift parameter space observed for SDSS quasars.  While our
sample covers more than an order of magnitude in \civ\ EQWs, we still
only have one source with a low \civ\ EQW and a large \civ\ blueshift
(see Figure \ref{fig:samplestats}).  This is Q2302 from the D09 sample,
and it is the object found to have the largest estimated \civ\ mass in
our sample.  Q2302 does appear as a significant outlier in the top panel
of Figure \ref{fig:masscomparefullsample} when its mass is estimated
using the \civ\ FWHM, again suggesting that FWHM-based \civ\ masses may
less reliable, but it is no larger an outlier than other sources without
large \civ\ blueshifts.  When the \civ\ mass is estimated with
$\sigma_l$, it also falls within the same, albeit much smaller, range of
scatter as the rest of the sample.  More low \civ\ EQW--large blueshift
objects should be specifically targeted for both SE mass comparisons and
RM studies to be able to address this concern further.

\subsection{C\,{\small{\bf IV}} Mass Scale Calibration}
\label{S_CIVmassScale}

\begin{figure}
\figurenum{9}
\epsscale{1.0}
\plotone{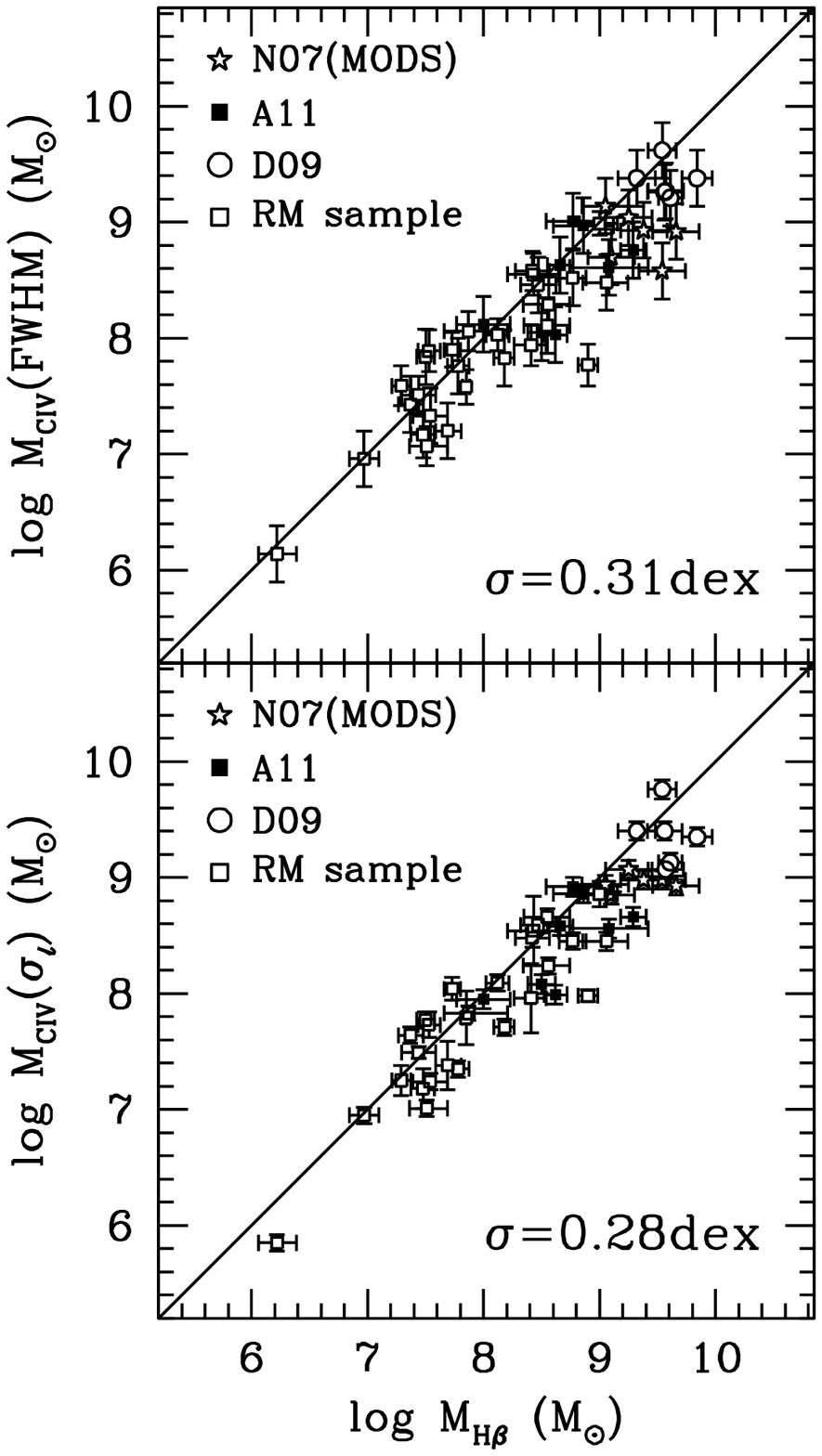}

\caption{Comparison between \Hbeta\ masses and \civ\ masses estimated
  from our complete sample of high quality data using the new \civ\ mass
  scaling relationships of P13.  The top (bottom) panel shows \civ\
  masses based on the FWHM ($\sigma_l$).  The solid line shows where the
  \civ\ and \Hbeta\ masses are equal.  The scatter, $\sigma$, quantified
  as the standard deviation about the mean of the sample of
  \civ-to-\Hbeta\ mass residuals, $\log M(\civm)-\log M({\rm H}\beta)$,
  is shown in the bottom right of each panel.}

\label{fig:newmassscale}
\end{figure}

The analysis and results presented above are based on the \civ\ mass
scale calibrated by VP06.  After completion and initial submission of
this current work, an updated calibration of the \civ\ mass scale became
available \citep[][hereafter P13]{Park13}.  The new calibration of P13
differs from that of VP06 in two main respects: P13 (1) incorporates the
most up-to-date database of high quality {\it HST} spectra of the
reverberation mapping sample (essentially the same sample we use here),
while excluding low quality spectra altogether, and (2) relaxes the
$M\propto V^2$ virial expectation, which improves the empirical
calibration of the FWHM-based \civ\ masses (see P13 for details).  We
have recalculated our \civ\ masses using the P13 \civ\ mass scaling
relationships to evaluate if the consistency between \civ\ and \Hbeta\
masses improves with these updated \civ\ mass scaling relations.  Figure
\ref{fig:newmassscale} shows the comparison of \civ\ and \Hbeta\ masses
that is equivalent to Figure \ref{fig:masscomparefullsample} but using
Equations (2) and (3) of P13.

The most significant difference between our previous results and those
utilizing these updated \civ\ mass scale calibrations is that the
scatter between the \civ\ FWHM-based masses and the \Hbeta\ masses is
significantly smaller.  This is a consequence of the $M_{\civm}(\rm
FWHM)$ dependence on $V^{0.56}$ instead of the virial expectation of
$V^2$, which helps to correct for line width dependent biases; it
effectively applies a line-width dependent scale factor to the masses
\citep[see][for similarly justified re-calibration of the \mgii\
FWHM-based SE mass scale]{Wang09, Rafiee&Hall11}.  For \civ\ this
corrects for the varying amounts of contamination by the non-variable
\civ\ emission component, which is a function of the \civ\ FWHM
\citep[see][]{Denney12}.  Such an empirical calibration may be the
answer for survey quality data, from which a measurement of the FWHM is
easier to make and more robust than $\sigma_l$ to $S/N$, but this type
of calibration is strongly sample dependent, particularly on the
distribution of \civ\ line shapes present in the calibration sample, and
the RM sample is relatively small and not yet representative of the
overall quasar population in this respect, so caution interpreting the
results of its application is necessary.  P13 find that a similar
relaxation of the virial dependence for $\sigma_l$-based \civ\ masses is
not warranted, and the P13 calibration for $\sigma_l$-based \civ\ masses
is very similar to the VP06 calibration.  We do not see any significant
improvement in the consistency between the $\sigma_l$-based \civ\ and
\Hbeta\ masses using the new calibration; the scatter was unaffected.
This demonstrates that $\sigma_l$ based masses continue to more
consistently reproduce \Hbeta-based BH masses with continued evidence
for a virial relation between the velocity dispersion of the gas
measured with $\sigma_l$ and the black hole mass.

A notable zero-point offset remains between the unity relation of the
\civ\ and \Hbeta\ mass and the distribution of points shown in Figure
\ref{fig:newmassscale}.  We attributed the offset observed in Figure
\ref{fig:masscomparefullsample} largely to prescriptional differences in
the line width measurements between our study and that of VP06.
However, we use a prescription for fitting the spectra and measuring the
line widths that is very similar to P13.  We have confirmed that this is
not the source of the offset here by comparing the line width
measurements from data we have in common with P13 (e.g., individual COS
spectra of the RM sample).  Instead, we traced the source of the offset
to the luminosity measurements. For the RM sample shared by both
studies, we find that the mass offset between the P13 \civ\ masses and
our own can be explained by (1) a minimal number of outliers resulting
from large luminosity differences because of the distance to these
objects (NGC4151, NGC4593, NGC4051, and NGC3873) adopted here and by
P13: P13 used the published redshifts to determine the luminosity, while
we used the best estimate of the distance measured from direct
measurement methods \citep[see][for details]{Bentz13}, (2) small, but
significant differences between L1350 and L1450 in another small subset
of targets with a relatively steeper UV continuum slope (despite direct
comparisons and a statement of general equivalence by both VP06 and
P13), and (3) intrinsic variability, which introduces luminosity
differences for individual objects because of the different method P13
used to combine multiple SE spectra of a single object compared to the
procedure we used here.  In all but one case (NGC4051), the former two
contributors lead to our \civ\ masses being lower than those estimated
by P13. The fact that the final contributor led to overall
systematically lower \civ\ masses is serendipitous.  In any case, the
resulting zero-point offsets in our masses, while not impacting our main
results here, underscores the importance for making direct measurements
of the \civ-emitting radius of the BLR with reverberation mapping, so
that \civ\ masses can be calibrated directly from a \civ\ $R-L$
relationship.  Such a calibration would be independent of the intrinsic
variability effects in the current calibration that are also the hardest
of the above contributors to mitigate.

\section{Summary and Concluding Remarks: Overall Impact of Data Quality}
\label{S_conclusion}

We have presented LBT/MODS1 spectra of the \civ\ emission line of six
high-redshift quasars that previously only had SDSS spectra.  We
expanded this small sample with 41 additional homogeneously re-analyzed,
high $S/N$ ($\geq$10 pixel$^{-1}$ in the continuum) spectra of the \civ\
emission-line region from the literature or public archives.  The most
significant improvements afforded by the increased data quality in the
MODS1 spectra over that of survey-quality data is the increased ability
to accurately define the intrinsic \civ\ emission-line profile and
underlying continuum and accurately identify the prevalence, location,
and strength of absorption.  With the advantage the data quality lends
to accurately characterizing the \civ\ line profile, SE \civ\ BH masses
can {\it reliably} be estimated from the virial relation using
high-quality data --- but only when using the line dispersion,
$\sigma_l$, to characterize the \civ\ emission-line width (Figure
\ref{fig:masscomparefullsample}).  The converse is true as well: SE
\civ\ masses can be reliably estimated with $\sigma_l$, {\it but only}
using high $S/N$ spectra (Figure \ref{fig:N07samplemasscompare}).  The
scatter, quantified as the standard deviation about the mean of the
residuals between the \Hbeta\ and $\sigma_l$-based \civ\ masses,
decreased by a factor of 2 (to 0.24\,dex) between results based on
survey-quality SDSS spectra and high $S/N$ MODS1 spectra for an, albeit,
small sample of AGNs.  Similarly, however, the same measure of scatter
in our full sample of 47 objects with high quality spectra was only
0.29\,dex.  Conversely, data quality had little impact on the scatter
between \Hbeta\ and FWHM-based \civ\ masses for our sample.  This
implies that obtaining high $S/N$ data cannot improve FWHM-based \civ\
masses the same way it can improve $\sigma_l$-based \civ\ masses.
Instead, it is possible that much of the scatter between FWHM-based
\civ\ and \Hbeta\ masses is due to the presence of the non-variable
\civ\ emission component described by \citet{Denney12}.  This component
significantly biases the BLR velocity measurement when the FWHM is used.
\civ\ masses based on the FWHM of the line profile and a virial mass
estimation should therefore {\it not be used}.

An alternative is to apply empirical corrections to survey quality,
FWHM-based mass estimates.  \citet{Denney12} provide an empirical
correction for FWHM-based \civ\ masses based on the ``shape'' of the
\civ\ line. \citet{Denney12} parameterized the line shape as the ratio
of the FWHM to $\sigma_l$ and found it to correlate with the
\civ-to-\Hbeta\ mass residuals.  However, since characterizing the line
shape this way requires a measurement of $\sigma_l$ anyway, one could
simply employ a $\sigma_l$-based \civ\ mass and avoid the need for a
correction altogether.  Measuring the shape from the kurtosis of the
line also produces a similar correlation with the \civ-to-\Hbeta\ mass
residuals, but regardless, accurately characterizing the line shape with
any parameterization is data-quality dependent, so this type of
correction is not likely to be as effective for survey quality data.
Alternately, the new empirically calibrated FWHM-based \civ\ mass
scaling relationship of P13 relaxes the virial assumption between the
FWHM and the BH mass in order to mitigate biases in the BH mass due to
the \civ\ FWHM measurement.  This reduces scatter between the \civ\ and
\Hbeta\ masses but removes much of the connection to the physical
assumptions behind virial BH mass estimators and will still have some
dependence on data quality --- both the current and P13 data sets were
high $S/N$, but \citet{Denney09a} describes reasonable expectations for
the effects the reduced $S/N$ of survey data may have on the dispersion
in the mass distribution.  Future work is planned to improve the
effectiveness of corrections for survey-quality FWHM-based masses
\citep[see also][]{Runnoe13}.  As with any correction of this type, the
potential for sample bias in the calibration is an issue, but such a
correction has the potential to improve \civ\ masses estimated from
survey-quality data at least somewhat.  

A11 also provide a correction to \civ\ SE masses based on the ratio of
UV-to-optical luminosities.  \citet{Benny&Netzer12} argue that such a
correlation is not broadly applicable because the rest frame optical
luminosity is rarely available for high-$z$ quasar samples.
Nonetheless, it is the implication behind the A11 correction that is of
the most interest for the \civ-based mass debate --- UV-to-optical
luminosity differences may be as much of a source of scatter in the
comparison of \civ-to-\Hbeta\ BH masses as the measurement of $\Delta
V$, implying that it is not specifically (or only) the \civ\ emission
line that is to blame for the observed discrepancies.

In the end, reliably determining \civ\ BH masses is of significant
interest to the larger astronomical community, since these masses are
essential for studying the cosmic evolution and growth of BHs and their
connection to galaxy evolution (e.g., feedback).  Even modest
improvements in precision and accuracy of BH mass estimates are a step
in the right direction and can add at least some additional constraints
to theoretical and model- or simulation-based studies of cosmological
evolution that depend on BH mass and growth rates.  Compared to all
other literature studies of this size and type, the relatively smaller
scatter observed here for the $\sigma_l$-based SE \civ\ masses, 0.28\,
dex, by using only high quality data implies that the best-achievable
precision in SE mass estimates is higher than previously believed.  In
particular, we fit the distribution of $\sigma_l$-based \civ\ masses
shown in Figure \ref{fig:newmassscale} with the IDL program
MPFITEXY\footnote{MPFITEXY \citep{Williams10} uses the MPFIT package of
  \citet{Markwardt09} combined with the procedure of \citet{Bedregal06}
  to estimate the intrinsic scatter.} to estimate the intrinsic scatter
of the \civ\ masses compared to the \Hbeta\ masses, by taking into
account the measurement and mass scale calibration uncertainties.  By
holding the slope fixed to a unity relation between the \civ\ and
\Hbeta\ mass, we find an estimate of the intrinsic scatter of these high
quality $\sigma_l$-based \civ\ masses to be 0.22\, dex; that for the
FWHM-based \civ\ masses is similar, 0.21\, dex.  This is on order the
observed scatter in the $R-L$ relationship on which the foundation of SE
mass estimates is built \citep[see][]{Bentz13}.  Future work is planned
to investigate additional applications of these results to better
understand the remaining scatter between \civ\ and \Hbeta\ masses.  The
potential for further reduction in the observed scatter between these
two quantities is promising, but we may be close to the limit of what is
possible for this method unless scatter in the $R-L$ relationship can be
further reduced.  Nonetheless, the application of such results to
studies of AGN physics and galaxy evolution will require larger samples
of high quality quasar spectra or development of new techniques that can
reliably characterize the BLR velocity from current survey quality
spectra.

\acknowledgements We would like to thank Matthias Dietrich for providing
\civ\ spectra for the objects from the D09 sample presented here.  KDD
acknowledges support from the People Programme (Marie Curie Actions) of
the European Union's Seventh Framework Programme FP7/2007-2013/ under
REA grant agreement no.\ 300553.  KDD, BMP, and MV acknowledge support
from grant HST-AR-12149 awarded by the Space Telescope Science
Institute, which is operated by the Association of Universities for
Research in Astronomy, Inc., for NASA, under contract NAS5-26555. RJA is
supported by an appointment to the NASA Postdoctoral Program at the Jet
Propulsion Laboratory, administered by Oak Ridge Associated Universities
through a contract with NASA. CSK is supported by NSF grant AST-1009756.
BMP and RWP are grateful for NSF support through grant AST-1008882 to
The Ohio State University.  The Dark Cosmology Centre is funded by the
Danish National Research Foundation.  This paper uses data taken with
the MODS spectrographs built with funding from NSF grant AST-9987045 and
the NSF Telescope System Instrumentation Program (TSIP), with additional
funds from the Ohio Board of Regents and the Ohio State University
Office of Research.  This work was based in part on observations made
with the Large Binocular Telescope.  The LBT is an international
collaboration among institutions in the United States, Italy and
Germany. The LBT Corporation partners are: the University of Arizona on
behalf of the Arizona university system; the Istituto Nazionale di
Astrofisica, Italy; the LBT Beteiligungsgesellschaft, Germany,
representing the Max Planck Society, the Astrophysical Institute
Potsdam, and Heidelberg University; the Ohio State University; and the
Research Corporation, on behalf of the University of Notre Dame, the
University of Minnesota, and the University of Virginia.



\begin{thebibliography}{98}
\expandafter\ifx\csname natexlab\endcsname\relax\def\natexlab#1{#1}\fi

\bibitem[{{Assef} {et~al.}(2011)}]{Assef11}
{Assef}, R.~J., {et~al.} 2011, \apj, 742, 93

\bibitem[{{Baskin} \& {Laor}(2005)}]{Baskin&Laor05}
{Baskin}, A., \& {Laor}, A. 2005, \mnras, 356, 1029

\bibitem[{{Bedregal} {et~al.}(2006){Bedregal}, {Arag{\'o}n-Salamanca}, \&
  {Merrifield}}]{Bedregal06}
{Bedregal}, A.~G., {Arag{\'o}n-Salamanca}, A., \& {Merrifield}, M.~R. 2006,
  \mnras, 373, 1125

\bibitem[{{Bentz} {et~al.}(2009{\natexlab{a}}){Bentz}, {Peterson}, {Netzer},
  {Pogge}, \& {Vestergaard}}]{Bentz09rl}
{Bentz}, M.~C., {Peterson}, B.~M., {Netzer}, H., {Pogge}, R.~W., \&
  {Vestergaard}, M. 2009{\natexlab{a}}, \apj, 697, 160

\bibitem[{{Bentz} {et~al.}(2006)}]{Bentz06b}
{Bentz}, M.~C., {et~al.} 2006, \apj, 651, 775

\bibitem[{{Bentz} {et~al.}(2007)}]{Bentz07}
---. 2007, \apj, 662, 205

\bibitem[{{Bentz} {et~al.}(2008)}]{Bentz08}
---. 2008, \apjl, 689, L21

\bibitem[{{Bentz} {et~al.}(2009{\natexlab{b}})}]{Bentz09lamp}
---. 2009{\natexlab{b}}, \apj, 705, 199

\bibitem[{{Bentz} {et~al.}(2010)}]{Bentz10b}
---. 2010, \apjl, 720, L46

\bibitem[{{Bentz} {et~al.}(2013)}]{Bentz13}
---. 2013, \apj, 767, 149

\bibitem[{{Blandford} \& {McKee}(1982)}]{Blandford82}
{Blandford}, R.~D., \& {McKee}, C.~F. 1982, \apj, 255, 419

\bibitem[{{Cappellari} {et~al.}(2002){Cappellari}, {Verolme}, {van der Marel},
  {Kleijn}, {Illingworth}, {Franx}, {Carollo}, \& {de Zeeuw}}]{Cappellari02}
{Cappellari}, M., {Verolme}, E.~K., {van der Marel}, R.~P., {Kleijn}, G.~A.~V.,
  {Illingworth}, G.~D., {Franx}, M., {Carollo}, C.~M., \& {de Zeeuw}, P.~T.
  2002, \apj, 578, 787

\bibitem[{{Cassinelli} \& {Castor}(1973)}]{Cassinelli73}
{Cassinelli}, J.~P., \& {Castor}, J.~I. 1973, \apj, 179, 189

\bibitem[{{Clavel} {et~al.}(1991)}]{Clavel91}
{Clavel}, J., {et~al.} 1991, \apj, 366, 64

\bibitem[{{Collin} {et~al.}(2006){Collin}, {Kawaguchi}, {Peterson}, \&
  {Vestergaard}}]{Collin06}
{Collin}, S., {Kawaguchi}, T., {Peterson}, B.~M., \& {Vestergaard}, M. 2006,
  \aap, 456, 75

\bibitem[{{Conroy} \& {White}(2013)}]{Conroy13}
{Conroy}, C., \& {White}, M. 2013, \apj, 762, 70

\bibitem[{{Davidson}(1972)}]{Davidson72}
{Davidson}, K. 1972, \apj, 171, 213

\bibitem[{{Denney}(2012)}]{Denney12}
{Denney}, K.~D. 2012, \apj, 759, 44

\bibitem[{{Denney} {et~al.}(2009{\natexlab{a}}){Denney}, {Peterson},
  {Dietrich}, {Vestergaard}, \& {Bentz}}]{Denney09a}
{Denney}, K.~D., {Peterson}, B.~M., {Dietrich}, M., {Vestergaard}, M., \&
  {Bentz}, M.~C. 2009{\natexlab{a}}, \apj, 692, 246

\bibitem[{{Denney} {et~al.}(2006)}]{Denney06}
{Denney}, K.~D., {et~al.} 2006, \apj, 653, 152

\bibitem[{{Denney} {et~al.}(2009{\natexlab{b}})}]{Denney09b}
---. 2009{\natexlab{b}}, \apj, 702, 1353

\bibitem[{{Denney} {et~al.}(2010)}]{Denney10}
---. 2010, \apj, 721, 715

\bibitem[{{Dietrich} {et~al.}(2009){Dietrich}, {Mathur}, {Grupe}, \&
  {Komossa}}]{Dietrich09}
{Dietrich}, M., {Mathur}, S., {Grupe}, D., \& {Komossa}, S. 2009, \apj, 696,
  1998

\bibitem[{{Dietrich} {et~al.}(1993)}]{Dietrich93}
{Dietrich}, M., {et~al.} 1993, \apj, 408, 416

\bibitem[{{Dietrich} {et~al.}(1998)}]{Dietrich98}
---. 1998, \apjs, 115, 185

\bibitem[{{Dietrich} {et~al.}(2012)}]{Dietrich12}
---. 2012, \apj, 757, 53

\bibitem[{{Ferrarese} \& {Ford}(2005)}]{Ferrarese05}
{Ferrarese}, L., \& {Ford}, H. 2005, Space Science Reviews, 116, 523

\bibitem[{{Fine} {et~al.}(2010){Fine}, {Croom}, {Bland-Hawthorn}, {Pimbblet},
  {Ross}, {Schneider}, \& {Shanks}}]{Fine10}
{Fine}, S., {Croom}, S.~M., {Bland-Hawthorn}, J., {Pimbblet}, K.~A., {Ross},
  N.~P., {Schneider}, D.~P., \& {Shanks}, T. 2010, \mnras, 409, 591

\bibitem[{{Gaskell}(1982)}]{Gaskell82}
{Gaskell}, C.~M. 1982, \apj, 263, 79

\bibitem[{{Gebhardt} {et~al.}(2003)}]{Gebhardt03}
{Gebhardt}, K., {et~al.} 2003, \apj, 583, 92

\bibitem[{{Gibson} {et~al.}(2009)}]{Gibson09}
{Gibson}, R.~R., {et~al.} 2009, \apj, 692, 758

\bibitem[{{Graham}(2008)}]{Graham08}
{Graham}, A.~W. 2008, \apj, 680, 143

\bibitem[{{Greene} {et~al.}(2010){Greene}, {Peng}, \& {Ludwig}}]{Greene10}
{Greene}, J.~E., {Peng}, C.~Y., \& {Ludwig}, R.~R. 2010, \apj, 709, 937

\bibitem[{{Grier} {et~al.}(2008)}]{Grier08}
{Grier}, C.~J., {et~al.} 2008, \apj, 688, 837

\bibitem[{{Grier} {et~al.}(2012)}]{Grier12b}
---. 2012, \apj, 755, 60

\bibitem[{{Grier} {et~al.}(2013{\natexlab{a}})}]{Grier13b}
---. 2013{\natexlab{a}}, ApJ, 773, 90

\bibitem[{{Grier} {et~al.}(2013{\natexlab{b}})}]{Grier13}
---. 2013{\natexlab{b}}, \apj, 764, 47

\bibitem[{{G{\"u}ltekin} {et~al.}(2009)}]{Gultekin09}
{G{\"u}ltekin}, K., {et~al.} 2009, \apj, 698, 198

\bibitem[{{Hamann} {et~al.}(2011){Hamann}, {Kanekar}, {Prochaska}, {Murphy},
  {Ellison}, {Malec}, {Milutinovic}, \& {Ubachs}}]{Hamann11}
{Hamann}, F., {Kanekar}, N., {Prochaska}, J.~X., {Murphy}, M.~T., {Ellison},
  S., {Malec}, A.~L., {Milutinovic}, N., \& {Ubachs}, W. 2011, \mnras, 410,
  1957

\bibitem[{{Ho} {et~al.}(2012){Ho}, {Goldoni}, {Dong}, {Greene}, \&
  {Ponti}}]{Ho12}
{Ho}, L.~C., {Goldoni}, P., {Dong}, X.-B., {Greene}, J.~E., \& {Ponti}, G.
  2012, \apj, 754, 11

\bibitem[{{Horne} {et~al.}(2004){Horne}, {Peterson}, {Collier}, \&
  {Netzer}}]{Horne04}
{Horne}, K., {Peterson}, B.~M., {Collier}, S.~J., \& {Netzer}, H. 2004, \pasp,
  116, 465

\bibitem[{{Kaspi} {et~al.}(2000){Kaspi}, {Smith}, {Netzer}, {Maoz}, {Jannuzi},
  \& {Giveon}}]{Kaspi00}
{Kaspi}, S., {Smith}, P.~S., {Netzer}, H., {Maoz}, D., {Jannuzi}, B.~T., \&
  {Giveon}, U. 2000, \apj, 533, 631

\bibitem[{{Kelly} {et~al.}(2010){Kelly}, {Vestergaard}, {Fan}, {Hopkins},
  {Hernquist}, \& {Siemiginowska}}]{Kelly10}
{Kelly}, B.~C., {Vestergaard}, M., {Fan}, X., {Hopkins}, P., {Hernquist}, L.,
  \& {Siemiginowska}, A. 2010, \apj, 719, 1315

\bibitem[{{Korista} {et~al.}(1995)}]{Korista95}
{Korista}, K.~T., {et~al.} 1995, \apjs, 97, 285

\bibitem[{{Krolik} \& {McKee}(1978)}]{Krolik&McKee78}
{Krolik}, J.~H., \& {McKee}, C.~F. 1978, \apjs, 37, 459

\bibitem[{{Leighly} \& {Moore}(2004)}]{Leighly04a}
{Leighly}, K.~M., \& {Moore}, J.~R. 2004, \apj, 611, 107

\bibitem[{{MacLeod} {et~al.}(2010)}]{Macleod10}
{MacLeod}, C.~L., {et~al.} 2010, \apj, 721, 1014

\bibitem[{{Markwardt}(2009)}]{Markwardt09}
{Markwardt}, C.~B. 2009, in Astronomical Society of the Pacific Conference
  Series, Vol. 411, Astronomical Data Analysis Software and Systems XVIII, ed.
  D.~A. {Bohlender}, D.~{Durand}, \& P.~{Dowler}, 251

\bibitem[{{McGill} {et~al.}(2008){McGill}, {Woo}, {Treu}, \&
  {Malkan}}]{McGill08}
{McGill}, K.~L., {Woo}, J.-H., {Treu}, T., \& {Malkan}, M.~A. 2008, \apj, 673,
  703

\bibitem[{{Metzroth} {et~al.}(2006){Metzroth}, {Onken}, \&
  {Peterson}}]{Metzroth06}
{Metzroth}, K.~G., {Onken}, C.~A., \& {Peterson}, B.~M. 2006, \apj, 647, 901

\bibitem[{{Netzer} {et~al.}(2007){Netzer}, {Lira}, {Trakhtenbrot}, {Shemmer},
  \& {Cury}}]{Netzer07}
{Netzer}, H., {Lira}, P., {Trakhtenbrot}, B., {Shemmer}, O., \& {Cury}, I.
  2007, \apj, 671, 1256

\bibitem[{{O'Brien} {et~al.}(1998)}]{OBrien98}
{O'Brien}, P.~T., {et~al.} 1998, \apj, 509, 163

\bibitem[{{Onken} {et~al.}(2004){Onken}, {Ferrarese}, {Merritt}, {Peterson},
  {Pogge}, {Vestergaard}, \& {Wandel}}]{Onken04}
{Onken}, C.~A., {Ferrarese}, L., {Merritt}, D., {Peterson}, B.~M., {Pogge},
  R.~W., {Vestergaard}, M., \& {Wandel}, A. 2004, \apj, 615, 645

\bibitem[{{Onken} \& {Peterson}(2002)}]{Onken02}
{Onken}, C.~A., \& {Peterson}, B.~M. 2002, \apj, 572, 746

\bibitem[{{Pancoast} {et~al.}(2012)}]{Pancoast12}
{Pancoast}, A., {et~al.} 2012, \apj, 754, 49

\bibitem[{{Park} {et~al.}(2012{\natexlab{a}}){Park}, {Kelly}, {Woo}, \&
  {Treu}}]{Park12b}
{Park}, D., {Kelly}, B.~C., {Woo}, J.-H., \& {Treu}, T. 2012{\natexlab{a}},
  \apjs, 203, 6

\bibitem[{{Park} {et~al.}(2013){Park}, {Woo}, {Denney}, \& {Shin}}]{Park13}
{Park}, D., {Woo}, J.-H., {Denney}, K.~D., \& {Shin}, J. 2013, \apj, 770, 87

\bibitem[{{Park} {et~al.}(2012{\natexlab{b}}){Park}, {Woo}, {Treu}, {Barth},
  {Bentz}, {Bennert}, {Canalizo}, {Filippenko}, {Gates}, {Greene}, {Malkan}, \&
  {Walsh}}]{Park12a}
{Park}, D., {Woo}, J.-H., {Treu}, T., {Barth}, A.~J., {Bentz}, M.~C.,
  {Bennert}, V.~N., {Canalizo}, G., {Filippenko}, A.~V., {Gates}, E., {Greene},
  J.~E., {Malkan}, M.~A., \& {Walsh}, J. 2012{\natexlab{b}}, \apj, 747, 30

\bibitem[{{Peterson}(1993)}]{Peterson93}
{Peterson}, B.~M. 1993, \pasp, 105, 247

\bibitem[{{Peterson} \& {Wandel}(1999)}]{Peterson99b}
{Peterson}, B.~M., \& {Wandel}, A. 1999, \apjl, 521, L95

\bibitem[{{Peterson} \& {Wandel}(2000)}]{Peterson00a}
---. 2000, \apjl, 540, L13

\bibitem[{{Peterson} {et~al.}(1998){Peterson}, {Wanders}, {Bertram}, {Hunley},
  {Pogge}, \& {Wagner}}]{Peterson98}
{Peterson}, B.~M., {Wanders}, I., {Bertram}, R., {Hunley}, J.~F., {Pogge},
  R.~W., \& {Wagner}, R.~M. 1998, \apj, 501, 82

\bibitem[{{Peterson} {et~al.}(2002)}]{Peterson02}
{Peterson}, B.~M., {et~al.} 2002, \apj, 581, 197

\bibitem[{{Peterson} {et~al.}(2004)}]{Peterson04}
---. 2004, \apj, 613, 682

\bibitem[{{Pogge} {et~al.}(2010)}]{Pogge10}
{Pogge}, R.~W., {et~al.} 2010, in Society of Photo-Optical Instrumentation
  Engineers (SPIE) Conference Series, Vol. 7735, Society of Photo-Optical
  Instrumentation Engineers (SPIE) Conference Series

\bibitem[{{Rafiee} \& {Hall}(2011{\natexlab{a}})}]{Rafiee&Hall11}
{Rafiee}, A., \& {Hall}, P.~B. 2011{\natexlab{a}}, \mnras, 415, 2932

\bibitem[{{Rafiee} \& {Hall}(2011{\natexlab{b}})}]{Rafiee11}
---. 2011{\natexlab{b}}, \apjs, 194, 42

\bibitem[{{Reichert} {et~al.}(1994)}]{Reichert94}
{Reichert}, G.~A., {et~al.} 1994, \apj, 425, 582

\bibitem[{{Richards} {et~al.}(2002){Richards}, {Vanden Berk}, {Reichard},
  {Hall}, {Schneider}, {SubbaRao}, {Thakar}, \& {York}}]{Richards02}
{Richards}, G.~T., {Vanden Berk}, D.~E., {Reichard}, T.~A., {Hall}, P.~B.,
  {Schneider}, D.~P., {SubbaRao}, M., {Thakar}, A.~R., \& {York}, D.~G. 2002,
  \aj, 124, 1

\bibitem[{{Richards} {et~al.}(2011)}]{Richards11}
{Richards}, G.~T., {et~al.} 2011, \aj, 141, 167

\bibitem[{{Rodriguez-Pascual} {et~al.}(1997)}]{Rodriguezpascual97}
{Rodriguez-Pascual}, P.~M., {et~al.} 1997, \apjs, 110, 9

\bibitem[{{Runnoe} {et~al.}(2013){Runnoe}, {Brotherton}, {Shang}, \&
  {DiPompeo}}]{Runnoe13}
{Runnoe}, J.~C., {Brotherton}, M.~S., {Shang}, Z., \& {DiPompeo}, M.~A. 2013,
  MNRAS, 434, 848

\bibitem[{{Santos-Lle{\' o}} {et~al.}(1997)}]{Santoslleo97}
{Santos-Lle{\' o}}, M., {et~al.} 1997, \apjs, 112, 271

\bibitem[{{Santos-Lle{\' o}} {et~al.}(2001)}]{Santoslleo01}
---. 2001, \aap, 369, 57

\bibitem[{{Schlafly} \& {Finkbeiner}(2011)}]{Schlafly11}
{Schlafly}, E.~F., \& {Finkbeiner}, D.~P. 2011, \apj, 737, 103

\bibitem[{{Shankar} {et~al.}(2013){Shankar}, {Weinberg}, \&
  {Miralda-Escud{\'e}}}]{Shankar13}
{Shankar}, F., {Weinberg}, D.~H., \& {Miralda-Escud{\'e}}, J. 2013, \mnras,
  428, 421

\bibitem[{{Shen} \& {Liu}(2012)}]{Shen12}
{Shen}, Y., \& {Liu}, X. 2012, \apj, 753, 125

\bibitem[{{Shen} {et~al.}(2011)}]{Shen11}
{Shen}, Y., {et~al.} 2011, \apjs, 194, 45

\bibitem[{{So{\l}tan}(1982)}]{Soltan82}
{So{\l}tan}, A. 1982, \mnras, 200, 115

\bibitem[{{Stirpe} {et~al.}(1994)}]{Stirpe94}
{Stirpe}, G.~M., {et~al.} 1994, \apj, 425, 609

\bibitem[{{Sulentic} {et~al.}(2007){Sulentic}, {Bachev}, {Marziani}, {Negrete},
  \& {Dultzin}}]{Sulentic07}
{Sulentic}, J.~W., {Bachev}, R., {Marziani}, P., {Negrete}, C.~A., \&
  {Dultzin}, D. 2007, \apj, 666, 757

\bibitem[{{Trakhtenbrot} \& {Netzer}(2012)}]{Benny&Netzer12}
{Trakhtenbrot}, B., \& {Netzer}, H. 2012, \mnras, 427, 3081

\bibitem[{{Trump} {et~al.}(2013){Trump}, {Hsu}, {Fang}, {Faber}, {Koo}, \&
  {Kocevski}}]{Trump13}
{Trump}, J.~R., {Hsu}, A.~D., {Fang}, J.~J., {Faber}, S.~M., {Koo}, D.~C., \&
  {Kocevski}, D.~D. 2013, \apj, 763, 133

\bibitem[{{Ulrich} \& {Horne}(1996)}]{Ulrich96}
{Ulrich}, M.-H., \& {Horne}, K. 1996, \mnras, 283, 748

\bibitem[{{van der Marel} \& {Franx}(1993)}]{vanderMarel93}
{van der Marel}, R.~P., \& {Franx}, M. 1993, \apj, 407, 525

\bibitem[{{Vanden Berk} {et~al.}(2004)}]{Vandenberk04}
{Vanden Berk}, D.~E., {et~al.} 2004, \apj, 601, 692

\bibitem[{{Vestergaard}(2003)}]{Vestergaard03}
{Vestergaard}, M. 2003, \apj, 599, 116

\bibitem[{{Vestergaard}(2004)}]{Vestergaard04}
---. 2004, \apj, 601, 676

\bibitem[{{Vestergaard} {et~al.}(2011){Vestergaard}, {Denney}, {Fan}, {Jensen},
  {Kelly}, {Osmer}, {Peterson}, \& {Tremonti}}]{Vestergaard11}
{Vestergaard}, M., {Denney}, K., {Fan}, X., {Jensen}, J.~J., {Kelly}, B.~C.,
  {Osmer}, P.~S., {Peterson}, B.~M., \& {Tremonti}, C.~A. 2011, in Narrow-Line
  Seyfert 1 Galaxies and their Place in the Universe

\bibitem[{{Vestergaard} \& {Osmer}(2009)}]{Vestergaard&Osmer09}
{Vestergaard}, M., \& {Osmer}, P.~S. 2009, \apj, 699, 800

\bibitem[{{Vestergaard} \& {Peterson}(2006)}]{Vestergaard06}
{Vestergaard}, M., \& {Peterson}, B.~M. 2006, \apj, 641, 689

\bibitem[{{Wanders} {et~al.}(1997)}]{Wanders97}
{Wanders}, I., {et~al.} 1997, \apjs, 113, 69

\bibitem[{{Wang} {et~al.}(2009){Wang}, {Dong}, {Wang}, {Ho}, {Yuan}, {Wang},
  {Zhang}, {Zhang}, \& {Zhou}}]{Wang09}
{Wang}, J., {Dong}, X., {Wang}, T., {Ho}, L.~C., {Yuan}, W., {Wang}, H.,
  {Zhang}, K., {Zhang}, S., \& {Zhou}, H. 2009, \apj, 707, 1334

\bibitem[{{Wild} {et~al.}(2008)}]{Wild08}
{Wild}, V., {et~al.} 2008, \mnras, 388, 227

\bibitem[{{Wilkes}(1984)}]{Wilkes84}
{Wilkes}, B.~J. 1984, \mnras, 207, 73

\bibitem[{{Williams} {et~al.}(2010){Williams}, {Bureau}, \&
  {Cappellari}}]{Williams10}
{Williams}, M.~J., {Bureau}, M., \& {Cappellari}, M. 2010, \mnras, 409, 1330

\bibitem[{{Woo} {et~al.}(2010)}]{Woo10}
{Woo}, J., {et~al.} 2010, \apj, 716, 269

\bibitem[{{Zu} {et~al.}(2011){Zu}, {Kochanek}, \& {Peterson}}]{Zu11}
{Zu}, Y., {Kochanek}, C.~S., \& {Peterson}, B.~M. 2011, \apj, 735, 80

\end{thebibliography}

\clearpage

\begin{deluxetable}{lcllllll}
\tablecolumns{8}
\tablewidth{0pt}
\tablecaption{Journal of Observations}
\tablehead{
\colhead{Object}&\colhead{z}&\colhead{UTC Date}&\colhead{Exposures}&\colhead{Channel}&\colhead{Grating}&\colhead{Seeing}&\colhead{Notes}}

\startdata
SDSSJ025438.37+002132.8  & 2.456     & 2011 Sept 28     & 3$\times$500s          & Blue            & G400L           & 0\farcs6       & Thin Cirrus\\
SDSSJ105511.99+020751.9  & 3.391     & 2012 Mar 23      & 3$\times$1600s         & Red             & G670L           & 0\farcs6     & Clear\\
SDSSJ115935.64+042420.0  & 3.451     & 2012 Apr 30      & 3$\times$1200s         & Red             & G670L           & 0\farcs6     & Clear\\
SDSSJ153725.36-014650.3  & 3.452     & 2012 Mar 24      & 3$\times$1800s         & Red             & G670L           & 0\farcs6     & Patchy clouds; wind\\
SDSSJ210258.22+002023.4  & 3.328     & 2011 Sept 28     & 3$\times$1300s         & Red             & G670L           & 0\farcs7     & Moderate Cirrus\\
SDSSJ210311.68-060059.4  & 3.336     & 2011 Sept 27     & 4$\times$250s          & Red             & G670L           & 0\farcs7     & Thin Cirrus
\enddata

\label{Tab_Observations}
\end{deluxetable}

\begin{deluxetable}{lcccccc}
\tablecolumns{7}
\tablewidth{0pt}
\tablecaption{N07 Sample Spectral Properties, Luminosities, and BH Masses}
\tabletypesize{\scriptsize}
\tablehead{
\colhead{Property}&\colhead{J0254\tablenotemark{a}}&\colhead{J1055}&\colhead{J1159\tablenotemark{a}}&\colhead{J1537}&\colhead{J2102}&\colhead{J2103\tablenotemark{a}}}

\startdata
{\bf SDSS Spectra}                       &              &              &              &               &              &              \\
$S/N$\tablenotemark{b}                   & 8.0          & 8.4          & 11.1         & 4.8           & 4.5          & 9.6          \\
\civ\ Obs.\ Line Boundaries (\AA)        & 5150--5600   & 6500--7100   & 6535--7200   & 6565--7030    & 6400--6950   & 6470--6930   \\
\civ\ FWHM(N07; km s$^{-1}$)             & 4753         & 5476         & 4160         & 5650          & 2355         & 4951         \\
\civ\ FWHM(this study; km s$^{-1}$)      & 3170$\pm$250 & 5680$\pm$620 & 3250$\pm$390 & 5980$\pm$1250 & 1340$\pm$840 & 2850$\pm$170 \\
\civ\ $\sigma_l$(this study; km s$^{-1}$)& 2930$\pm$140 & 4100$\pm$130 & 4110$\pm$110 & 3910$\pm$720  & 2260$\pm$280 & 2960$\pm$130 \\
log\,$\lambda L_{\lambda}$(1450\AA)(erg s$^{-1}$)      & 45.93        & 46.24        & 46.43        & 46.44         & 45.86        & 46.24        \\     
log\,$M_{\civm}$(FWHM)(M$_{\odot}$)      & 8.69$\pm$0.08& 9.36$\pm$0.10& 8.97$\pm$0.11& 9.51$\pm$0.19 & 7.90$\pm$0.55& 8.76$\pm$0.07\\     
\vspace{0.05in}
log\,$M_{\civm}$($\sigma_l$)(M$_{\odot}$)& 8.69$\pm$0.06& 9.14$\pm$0.05& 9.25$\pm$0.05& 9.21$\pm$0.17 & 8.42$\pm$0.12& 8.86$\pm$0.04\\
\hline
{\bf MODS Spectra}                       &              &              &              &               &              &              \\
$S/N$\tablenotemark{b}                   & 17.8         & 80.4         & 50.5         & 51.9          & 31.3         & 47.3         \\
\civ\ Line Boundaries (\AA)              & 5100--5600   & 6515--7070  & 6665--7180   & 6585--7180    & 6300--6950   & 6470--6930   \\     
\civ\ FWHM (km s$^{-1}$)                 & 2440$\pm$100 & 4980$\pm$180 & 2130$\pm$80  & 4910$\pm$170  & 1750$\pm$70  & 3120$\pm$70  \\
\civ\ $\sigma_l$ (km s$^{-1}$)           & 3870$\pm$40  & 4170$\pm$30  & 3350$\pm$40  & 3200$\pm$50   & 4660$\pm$80  & 3560$\pm$30  \\
log\,$M_{\civm}$(FWHM)(M$_{\odot}$)      & 8.46$\pm$0.06& 9.24$\pm$0.05& 8.61$\pm$0.05& 9.34$\pm$0.05 & 8.13$\pm$0.06& 8.84$\pm$0.05\\     
\vspace{0.05in}
log\,$M_{\civm}$($\sigma_l$)(M$_{\odot}$)& 8.93$\pm$0.04& 9.16$\pm$0.04& 9.07$\pm$0.04& 9.03$\pm$0.05 & 9.05$\pm$0.05& 9.02$\pm$0.04\\   
\hline
{\bf Gemini Spectra; N07}                &              &              &              &               &              &              \\ 
\Hbeta\ FWHM(km s$^{-1}$)                & 4164         & 5424         & 5557         & 3656          & 7198         & 6075         \\
log\,$\lambda L_{\lambda}$(5100\AA)(erg s$^{-1}$)      & 45.85        & 45.85        & 45.92        & 45.98         & 45.79        & 46.30        \\
log\,$M_{\Hbeta}$(FWHM)(M$_{\odot}$)     & 9.162        & 9.294        & 9.460        & 9.133         & 9.599        & 9.785                     
\enddata

\tablenotetext{a}{The \civ\ profile of this object was observed to have
  absorption across the line peak.}  
\tablenotetext{b}{$S/N$ was measured per resolution element in the
  continuum near rest frame 1700\AA.}

\label{Tab_N07Widths}
\end{deluxetable}

\clearpage
\begin{landscape}
\begin{deluxetable}{llccccccccc} 
\tablewidth{0pt}
\tablecaption{D09 Sample Spectral Parameters and Masses}
\tabletypesize{\tiny}
\tablecolumns{11}
\tablehead{
\colhead{ D09} &
\colhead{ NED\tablenotemark{a}} &
\colhead{ } &
\colhead{ Res.} &
\colhead{ \civ\ Rest Frame} &
\colhead{ log\,$\lambda L_{\lambda}$(1350\AA)} &
\colhead{ FWHM(C{\sc IV})} &
\colhead{ $\sigma_l$(C{\sc IV})} &
\colhead{ log\,$M_{\civm}$(FWHM)} &
\colhead{ log\,$M_{\civm}$($\sigma_l$)} &
\colhead{ log\,$M_{\Hbeta}$} \\
\colhead{ Object ID} &
\colhead{ Object ID} &
\colhead{ $z$} &
\colhead{ (\AA)} &
\colhead{ Boundaries} &
\colhead{ (erg s$^{-1}$)} &
\colhead{ (km s$^{-1}$)} &
\colhead{ (km s$^{-1}$)} &
\colhead{ (M$_{\odot}$)} &
\colhead{ (M$_{\odot}$)} &
\colhead{ (M$_{\odot}$)} \\
\colhead{(1)} &
\colhead{(2)} &
\colhead{(3)} &
\colhead{(4)} &
\colhead{(5)} &
\colhead{(6)} &
\colhead{(7)} &
\colhead{(8)} &
\colhead{(9)} &
\colhead{(10)} &
\colhead{(11)} 
}
\startdata
Q\,0150-202  & [HB89]\,0150-202 & 2.147 & 2.8 &1465--1597&46.970& 4230$\pm$310 & 3790$\pm$100&  9.49$\pm$0.08&9.46$\pm$0.05&9.84$\pm$0.13\\
Q\,2116-4439 & LBQS\,2116-4439  & 1.504 & 2.8 &1477--1635&46.712& 7540$\pm$140 & 4660$\pm$50 &  9.85$\pm$0.05&9.50$\pm$0.04&9.32$\pm$0.16\\
Q\,2154-2005 & LBQS\,2154-2005  & 2.042 & 2.8 &1490--1625&46.681& 5030$\pm$190 & 3250$\pm$70 &  9.49$\pm$0.05&9.17$\pm$0.05&9.57$\pm$0.15\\
Q\,2209-1842 & LBQS\,2209-1842  & 2.098 & 2.8 &1480--1610&46.808& 3020$\pm$100 & 3230$\pm$50 &  9.11$\pm$0.05&9.24$\pm$0.05&9.61$\pm$0.10\\
Q\,2230+0232 & LBQS\,2230+0232  & 2.215 & 2.8 &1470--1640&46.724& 4540$\pm$160 & 4590$\pm$60 &  9.42$\pm$0.05&9.50$\pm$0.05&9.56$\pm$0.15\\
Q\,2302+0255 & [HB89]\,2302+029 & 1.062 & 2.8 &1440--1610&46.915& 12940$\pm$690& 6270$\pm$140& 10.43$\pm$0.06&9.87$\pm$0.05&9.54$\pm$0.12  
\enddata

\tablenotetext{a}{The NASA/IPAC Extragalactic Database (NED) is operated
  by the Jet Propulsion Laboratory, California Institute of Technology,
  under contract with the National Aeronautics and Space
  Administration.}  

\label{Tab_D09widths}
\end{deluxetable}
\clearpage
\end{landscape}

\LongTables
\begin{deluxetable}{lccccl} 
\tablecolumns{6}
\tablewidth{0pt}
\tablecaption{Reverberation Sample \Hbeta\ Rest-frame Lags, Line Widths, and Masses\label{Tab_Hbeta_masses}}
\tabletypesize{\scriptsize}
\tablehead{
\colhead{ } &
\colhead{ } &
\colhead{ $\tau_{\rm cent}$} &
\colhead{ $\sigma_l$(RMS)} &
\colhead{ log\,$M_{\rm RM}$(\Hbeta)\tablenotemark{a,b}} &
\colhead{ } \\
\colhead{ Object} &
\colhead{ $z$} &
\colhead{ Restframe} &
\colhead{ (km s$^{-1}$)} &
\colhead{ (M$_{\odot}$)} &
\colhead{ (Ref.)} \\
\colhead{(1)} &
\colhead{(2)} &
\colhead{(3)} &
\colhead{(4)} &
\colhead{(5)} &
\colhead{(6)} 
}
\startdata
Mrk\,335   & 0.02578 & 16.80$^{+4.80}_{-4.20}$    &  917$\pm$52  & 7.16$^{+0.16}_{-0.15}$ & 1,25        \\ 
	   & 0.02578 & 12.50$^{+6.60}_{-5.50}$    &  948$\pm$113 & 7.06$^{+0.27}_{-0.24}$ & 1,25        \\ 
	   & 0.02578 & 14.30$^{+0.70}_{-0.70}$    & 1293$\pm$64	 & 7.39$^{+0.10}_{-0.10}$ & 2           \\ 
	   & 	     &      		          &		 & \boldmath$7.29^{+0.08}_{-0.08}$ &    \\
PG0026+129 & 0.14200 & 111.00$^{+24.10}_{-28.30}$ & 1773$\pm$285 & 8.55$^{+0.19}_{-0.20}$ & 3,25        \\ 
PG0052+251 & 0.15500 & 89.80$^{+24.50}_{-24.10}$  & 1783$\pm$86	 & 8.47$^{+0.15}_{-0.15}$ & 3,25        \\ 
Fairall 9  & 0.04702 & 17.40$^{+3.20}_{-4.30}$    & 3787$\pm$197 & 8.41$^{+0.13}_{-0.15}$ & 4,5,25      \\ 
Mrk\,590   & 0.02638 & 20.70$^{+3.50}_{-2.70}$    &  789$\pm$74	 & 7.12$^{+0.14}_{-0.13}$ & 1,25        \\ 
	   & 0.02638 & 14.00$^{+8.50}_{-8.80}$    & 1935$\pm$52	 & 7.73$^{+0.28}_{-0.29}$ & 1,25        \\ 
	   & 0.02638 & 29.20$^{+4.90}_{-5.00}$    & 1251$\pm$72	 & 7.67$^{+0.13}_{-0.13}$ & 1,25        \\ 
	   & 0.02638 & 28.80$^{+3.60}_{-4.20}$    & 1201$\pm$130 & 7.63$^{+0.14}_{-0.14}$ & 1,25        \\ 
	   & 	     &      			  &	     	 & \boldmath$7.50^{+0.08}_{-0.08}$ &    \\
3C\,120	   & 0.03301 & 38.10$^{+21.30}_{-15.30}$  & 1166$\pm$50	 & 7.73$^{+0.26}_{-0.20}$ & 1,25        \\ 
	   & 0.03301 & 25.90$^{+2.30}_{-2.30}$    & 1514$\pm$65	 & 7.79$^{+0.10}_{-0.10}$ & 2           \\ 
	   & 	     &      			  &	     	 & \boldmath$7.78^{+0.10}_{-0.09}$ &    \\
Akn\,120   & 0.03230 & 47.10$^{+8.30}_{-12.40}$   & 1959$\pm$109 & 8.27$^{+0.13}_{-0.15}$ & 1,25        \\ 
	   & 0.03230 & 37.10$^{+4.80}_{-5.40}$    & 1884$\pm$48	 & 8.13$^{+0.11}_{-0.11}$ & 1,25        \\ 
	   & 	     &      			  &	     	 & \boldmath$8.18^{+0.08}_{-0.09}$ &    \\
PG0804+761 & 0.10000 & 146.90$^{+18.80}_{-18.90}$ & 1971$\pm$105 & 8.77$^{+0.12}_{-0.12}$ & 3,25        \\ 
PG0953+414 & 0.23410 & 150.10$^{+21.60}_{-22.60}$ & 1306$\pm$144 & 8.42$^{+0.15}_{-0.15}$ & 3,25        \\ 
NGC\,3516  & 0.00884 & 11.68$^{+1.02}_{-1.53}$    & 1591$\pm$10	 & 7.48$^{+0.10}_{-0.11}$ & 6           \\ 
NGC\,3783  & 0.00584\tablenotemark{c}& 10.20$^{+3.30}_{-2.30}$    & 1753$\pm$141 & 7.51$^{+0.18}_{-0.15}$ & 7,8,9,25    \\ 
NGC\,4051  & 0.00397\tablenotemark{c}& 1.87$^{+0.54}_{-0.50}$     &  927$\pm$64	 & 6.22$^{+0.17}_{-0.16}$ & 10          \\ 
NGC\,4151  & 0.0026\tablenotemark{c} & 6.59$^{+1.12}_{-0.76}$     & 2680$\pm$64	 & 7.69$^{+0.12}_{-0.11}$ & 11          \\ 
PG1226+023 & 0.15834 & 306.80$^{+68.50}_{-90.90}$ & 1777$\pm$150 & 9.00$^{+0.15}_{-0.17}$ & 3,25        \\ 
PG1229+204 & 0.06301 & 37.80$^{+27.60}_{-15.30}$  & 1385$\pm$111 & 7.87$^{+0.34}_{-0.21}$ & 3,25        \\ 
NGC\,4593  & 0.00865\tablenotemark{c}& 3.73$^{+0.75}_{-0.75}$     & 1561$\pm$55	 & 6.97$^{+0.13}_{-0.13}$ & 12          \\ 
PG1307+085 & 0.15500 & 105.60$^{+36.00}_{-46.60}$ & 1820$\pm$122 & 8.56$^{+0.18}_{-0.22}$ & 3,25        \\ 
Mrk\,279   & 0.03045 & 16.70$^{+3.90}_{-3.90}$    & 1420$\pm$96	 & 7.54$^{+0.15}_{-0.15}$ & 13,25       \\ 
NGC\,5548  & 0.01717 & 19.70$^{+1.50}_{-1.50}$    & 1687$\pm$56	 & 7.76$^{+0.10}_{-0.10}$ & 14,15,16,25 \\ 
	   & 0.01717 & 18.60$^{+2.10}_{-2.30}$    & 1882$\pm$83	 & 7.83$^{+0.11}_{-0.11}$ & 14,25       \\ 
	   & 0.01717 & 15.90$^{+2.90}_{-2.50}$    & 2075$\pm$81	 & 7.85$^{+0.12}_{-0.12}$ & 14,25       \\ 
	   & 0.01717 & 11.00$^{+1.90}_{-2.00}$    & 2264$\pm$88	 & 7.76$^{+0.12}_{-0.12}$ & 14,25       \\ 
	   & 0.01717 & 13.00$^{+1.60}_{-1.40}$    & 1909$\pm$129 & 7.69$^{+0.12}_{-0.12}$ & 14,17,25    \\ 
 	   & 0.01717 & 13.40$^{+3.80}_{-4.30}$    & 2895$\pm$114 & 8.06$^{+0.16}_{-0.17}$ & 14,25       \\ 
 	   & 0.01717 & 21.70$^{+2.60}_{-2.60}$    & 2247$\pm$134 & 8.05$^{+0.12}_{-0.12}$ & 14,25       \\ 
 	   & 0.01717 & 16.40$^{+1.20}_{-1.10}$    & 2026$\pm$68	 & 7.84$^{+0.10}_{-0.10}$ & 14,25       \\ 
 	   & 0.01717 & 17.50$^{+2.00}_{-1.60}$    & 1923$\pm$62	 & 7.82$^{+0.11}_{-0.10}$ & 14,25       \\ 
 	   & 0.01717 & 26.50$^{+4.30}_{-2.20}$    & 1732$\pm$76	 & 7.91$^{+0.12}_{-0.10}$ & 14,25       \\ 
 	   & 0.01717 & 24.80$^{+3.20}_{-3.00}$    & 1980$\pm$30	 & 8.00$^{+0.11}_{-0.11}$ & 14,25       \\ 
 	   & 0.01717 & 6.50$^{+5.70}_{-3.70}$     & 1969$\pm$48	 & 7.41$^{+0.39}_{-0.26}$ & 14,25       \\ 
 	   & 0.01717 & 14.30$^{+5.90}_{-7.30}$    & 2173$\pm$89	 & 7.84$^{+0.20}_{-0.24}$ & 14,25       \\ 
 	   & 0.01717 & 6.30$^{+2.60}_{-2.30}$     & 3210$\pm$642 & 7.82$^{+0.27}_{-0.25}$ & 18          \\ 
 	   & 0.01717 & 12.40$^{+2.74}_{-3.85}$    & 1822$\pm$35	 & 7.63$^{+0.13}_{-0.16}$ & 6           \\ 
 	   & 0.01717 & 4.18$^{+0.86}_{-1.30}$     & 4270$\pm$292 & 7.89$^{+0.14}_{-0.17}$ & 19          \\ 
	   & 	     &      			  &	     	 & \boldmath$7.85^{+0.03}_{-0.03}$ &    \\
PG1426+015 & 0.08647 & 95.00$^{+29.90}_{-37.10}$  & 3442$\pm$308 & 9.06$^{+0.18}_{-0.21}$ & 3,25        \\ 
Mrk\,817   & 0.03145 & 19.00$^{+3.90}_{-3.70}$    & 1392$\pm$78	 & 7.58$^{+0.14}_{-0.13}$ & 1,25        \\ 
	   & 0.03145 & 15.30$^{+3.70}_{-3.50}$    & 1971$\pm$96	 & 7.79$^{+0.14}_{-0.14}$ & 1,25        \\ 
	   & 0.03145 & 33.60$^{+6.50}_{-7.60}$    & 1729$\pm$158 & 8.01$^{+0.15}_{-0.16}$ & 1,25        \\ 
	   & 0.03145 & 14.04$^{+3.41}_{-3.47}$    & 2025$\pm$5\tablenotemark{d} & 7.77$^{+0.14}_{-0.14}$ & 6 \\ 
	   & 	     &      			  &		 & \boldmath$7.78^{+0.07}_{-0.07}$ &    \\
Mrk\,290   & 0.02958 & 8.72$^{+1.21}_{-1.02}$     & 1609$\pm$47	 & 7.37$^{+0.11}_{-0.11}$ & 6           \\ 
PG1613+658 & 0.12900 & 40.10$^{+15.00}_{-15.20}$  & 2547$\pm$342 & 8.43$^{+0.22}_{-0.22}$ & 3,25        \\ 
3C\,390.3  & 0.05610 & 23.60$^{+6.20}_{-6.70}$    & 3105$\pm$81	 & 8.37$^{+0.15}_{-0.15}$ & 20,21,25    \\ 
	   & 0.05610 & 46.40$^{+3.60}_{-3.20}$    & 5455$\pm$278 & 9.15$^{+0.11}_{-0.10}$ & 22          \\ 
	   & 	     &      			  &		 & \boldmath$8.90^{+0.09}_{-0.09}$ &    \\
Mrk\,509   & 0.03440 & 79.60$^{+6.10}_{-5.40}$    & 1276$\pm$28	 & 8.12$^{+0.10}_{-0.10}$ & 1,25        \\ 
PG2130+099 & 0.06298 & 22.90$^{+4.70}_{-4.60}$    & 1246$\pm$222 & 7.56$^{+0.20}_{-0.20}$ & 23          \\ 
           & 0.06298 & 9.60$^{+1.20}_{-1.20}$     & 1825$\pm$65	 & 7.52$^{+0.11}_{-0.11}$ & 2           \\ 
	   & 	     &      			  &		 & \boldmath$7.53^{+0.10}_{-0.10}$ &    \\
NGC\,7469  & 0.01632 & 16.50$^{+2.90}_{-2.90}$    & 1274$\pm$126 & 7.44$^{+0.15}_{-0.15}$ & 24	
\enddata

\tablenotetext{a}{Assumes $\log f=0.72\pm0.09$ \citep{Woo10} except for
  the one season of Mrk817 observations when the line width was measured
  from the mean spectrum; here $\log f=0.59$ \citep{Collin06}.}
\tablenotetext{b}{Values in bold are the weighted mean; see Section
  \ref{S_widthsMasses} for details.}
\tablenotetext{c}{This redshift has been modified to reflect the most
  probable true distance \citep[see][]{Bentz13}.}
\tablenotetext{d}{This line width was measured in the mean, not the rms
  spectrum.  See \citet{Denney10} for details.}
\tablerefs{(1)\citet{Peterson98}; (2)\citet{Grier12b};
  (3)\citet{Kaspi00}; (4)\citet{Santoslleo97};
  (5)\citet{Rodriguezpascual97}; (6)\citet{Denney10};
  (7)\citet{Stirpe94}; (8)\citet{Onken02}; (9)\citet{Reichert94};
  (10)\citet{Denney09b}; (11)\citet{Bentz06b}; (12)\citet{Denney06};
  (13)\citet{Santoslleo01}; (14)\citet[][and references
  therein]{Peterson02}; (15)\citet{Dietrich93}; (16)\citet{Clavel91};
  (17)\citet{Korista95}; (18)\citet{Bentz07}; (19)\citet{Bentz09lamp};
  (20)\citet{Dietrich98}; (21)\citet{OBrien98}; (22)\citet{Dietrich12};
  (23)\citet{Grier08}; (24) Peterson et al.\ (2013, in prep.);
  (25)Reanalyzed by \citet{Peterson04}.}

\end{deluxetable}
\clearpage

\LongTables
\begin{landscape}
\begin{deluxetable}{lcllccccccc} 
\tablewidth{0pt}
\tablecaption{Reverberation Sample \civ\ Spectral Parameters and Masses\label{Tab_RMwidths}}
\tabletypesize{\tiny}
\tablecolumns{11}
\tablehead{
\colhead{ } &
\colhead{ } &
\colhead{ Date} &
\colhead{ Telescope/} &
\colhead{ Resolution\tablenotemark{a}} &
\colhead{ \civ\ Obs.\ Frame} &
\colhead{ log\,$\lambda L_{\lambda}$(1450\AA)} &
\colhead{ FWHM(C{\sc IV})} &
\colhead{ $\sigma_l$(C{\sc IV})} &
\colhead{ log\,$M_{\civm}$(FWHM)\tablenotemark{b}} &
\colhead{ log\,$M_{\civm}$($\sigma_l$)\tablenotemark{b}} \\
\colhead{ Object} &
\colhead{ $z$} &
\colhead{ Observed} &
\colhead{ Instrument} &
\colhead{ (\AA)} &
\colhead{ Boundaries} &
\colhead{ (erg s$^{-1}$)} &
\colhead{ (km s$^{-1}$)} &
\colhead{ (km s$^{-1}$)} &
\colhead{ (M$_{\odot}$)} &
\colhead{ (M$_{\odot}$)} \\
\colhead{(1)} &
\colhead{(2)} &
\colhead{(3)} &
\colhead{(4)} &
\colhead{(5)} &
\colhead{(6)} &
\colhead{(7)} &
\colhead{(8)} &
\colhead{(9)} &
\colhead{(10)} &
\colhead{(11)} 
}
\startdata
%
Mrk\,335   & 0.02578 & 1994 Dec 16	& \HST/FOS    & 1.40 & 1553--1625 & 44.129$\pm$0.002 & 2000$\pm$60  & 2020$\pm$20  & 7.33$\pm$0.03 & 7.41$\pm$0.01 \\
	   & 0.02578 & 2009 Oct 31	& \HST/COS    & 0.21 & 1550--1635 & 43.850$\pm$0.001 & 1730$\pm$30  & 1880$\pm$10  & 7.06$\pm$0.02 & 7.20$\pm$0.01 \\
	   & 0.02578 & 2010 Feb 8	& \HST/COS\tablenotemark{e} & 0.21 & 1550--1635 & 43.850$\pm$0.001 & 1690$\pm$30  & 1860$\pm$10  & 7.04$\pm$0.02 & 7.19$\pm$0.01 \\
	   &	     &			& 	     &      & 	         &                  & 	           &     & \boldmath$7.10\pm0.16$ & \boldmath$7.25\pm0.12$ \\
PG0026+129 & 0.14200 & 1994 Nov 27	& \HST/FOS\tablenotemark{e} & 1.40 & 1660--1870 & 45.007$\pm$0.004 & 1540$\pm$110 & 5310$\pm$130 & 7.57$\pm$0.06 & 8.71$\pm$0.02 \\
PG0052+251 & 0.15500 & 1993 July 22	& \HST/FOS\tablenotemark{e} & 2.20 & 1690--1880 & 45.064$\pm$0.006 & 5710$\pm$370 & 4740$\pm$80  & 8.74$\pm$0.06 & 8.65$\pm$0.02 \\
Fairall 9  & 0.04702 & 1994 RM\tablenotemark{d}	& \IUE/SWP    & 6.00 & 1535--1670 & 44.479$\pm$0.003 & 2830$\pm$60  & 4080$\pm$40  & 7.82$\pm$0.02 & 8.21$\pm$0.01	\\
	   & 0.04702 & 1993 Jan 22	& \HST/FOS\tablenotemark{e} & 2.20 & 1560--1680 & 44.360$\pm$0.004 & 2510$\pm$70  & 2710$\pm$30  & 7.65$\pm$0.03 & 7.79$\pm$0.01	\\
	   & 	     &   	        & 	     & 	    & 	         & 	            & 	  	   & 	 & \boldmath$7.75\pm0.12$ & \boldmath$8.01\pm0.30$	\\
Mrk\,590   & 0.02638 & 1991 Jan 14	& \IUE/SWP\tablenotemark{e} & 6.00 & 1520--1670 & 43.961$\pm$0.009 & 4690$\pm$230 & 3480$\pm$80  & 7.98$\pm$0.04 & 7.79$\pm$0.02	\\
3C\,120	   & 0.03301 & 1993 Aug 25	& \IUE/SWP\tablenotemark{e} & 6.00 & 1570--1645 & 43.959$\pm$0.008 & 3450$\pm$270 & 2150$\pm$60  & 7.71$\pm$0.07 & 7.37$\pm$0.03	\\
Akn\,120   & 0.03230 & 1995 Jul 29	& \HST/FOS\tablenotemark{e} & 1.40 & 1545--1655 & 44.038$\pm$0.004 & 3900$\pm$130 & 3090$\pm$30  & 7.86$\pm$0.03 & 7.73$\pm$0.01	\\
PG0804+761 & 0.10000 & 2010 Jun 12	& \HST/COS\tablenotemark{e} & 2.10 & 1640--1770 & 45.396$\pm$0.001 & 3630$\pm$130 & 3320$\pm$70  & 8.52$\pm$0.03 & 8.51$\pm$0.02	\\
PG0953+414 & 0.23410 & 1991 Jun 17	& \HST/FOS    & 1.50 & 1850--1980 & 45.587$\pm$0.003 & 2860$\pm$100 & 3230$\pm$40  & 8.41$\pm$0.03 & 8.59$\pm$0.01	\\
 	   & 0.23410 & 2001 Jan 21	& \HST/STIS\tablenotemark{e}& 3.14 & 1840--1980 & 45.650$\pm$0.005 & 2900$\pm$120 & 2830$\pm$80  & 8.46$\pm$0.04 & 8.51$\pm$0.03	\\
 	   & 	     &   	        & 	     & 	    & 	         & 	            & 	  	   & 	 & \boldmath$8.43\pm0.04$ & \boldmath$8.57\pm0.06$	\\
NGC\,3516\tablenotemark{f}  & 0.00884 & 1995 Dec 30	& \HST/FOS    & 1.40 & 1475--1620 & 42.671$\pm$0.003 & 3050$\pm$90  & 3460$\pm$40  & 6.92$\pm$0.03 & 7.10$\pm$0.01	\\
 	   & 0.00884 & 1996 Feb 21	& \HST/FOS    & 1.40 & 1475--1620 & 43.038$\pm$0.002 & 4750$\pm$160 & 3490$\pm$50  & 7.50$\pm$0.03 & 7.31$\pm$0.02	\\
 	   & 0.00884 & 1996 Apr 13	& \HST/FOS    & 1.40 & 1475--1620 & 42.990$\pm$0.002 & 3850$\pm$150 & 3250$\pm$40  & 7.30$\pm$0.04 & 7.22$\pm$0.01	\\
 	   & 0.00884 & 1996 Aug 14	& \HST/FOS    & 1.40 & 1475--1620 & 42.869$\pm$0.002 & 4130$\pm$90  & 3380$\pm$40  & 7.29$\pm$0.02 & 7.19$\pm$0.01	\\
 	   & 0.00884 & 1996 Nov 28	& \HST/FOS    & 1.40 & 1475--1630 & 42.345$\pm$0.005 & 2990$\pm$100 & 3170$\pm$40  & 6.73$\pm$0.03 & 6.85$\pm$0.02	\\
 	   & 0.00884 & 1998 Apr 13	& \HST/STIS\tablenotemark{e}& 1.20 & 1475--1640 & 42.659$\pm$0.002 & 4890$\pm$80  & 4480$\pm$20  & 7.33$\pm$0.02 & 7.32$\pm$0.01	\\
 	   & 	     &   	        & 	     & 	    & 	         & 	            & 	  	   & 	 & \boldmath$7.22\pm0.29$ & \boldmath$7.18\pm0.17$	\\
NGC\,3783  & 0.00584\tablenotemark{c}& 1992 RM\tablenotemark{d}	& \IUE/SWP    & 6.00 & 1520--1610 & 42.702$\pm$0.003 & 2900$\pm$60  & 2860$\pm$20  & 6.90$\pm$0.02 & 6.95$\pm$0.01	\\
 	   & 0.00584\tablenotemark{c}& 1992 Jul 22	& \HST/FOS\tablenotemark{e} & 1.95 & 1520--1630 & 42.854$\pm$0.002 & 2270$\pm$80  & 2830$\pm$20  & 6.76$\pm$0.03 & 7.03$\pm$0.01	\\
 	   & 	     &   	        & 	     & 	    & 	         & 	            & 	  	   & 	 & \boldmath$6.86\pm0.10$ & \boldmath$6.99\pm0.05$	\\
NGC\,4051\tablenotemark{f}  & 0.00397\tablenotemark{c}& 2000 Mar 25	& \HST/STIS\tablenotemark{e}& 0.51 & 1520--1585 & 41.796$\pm$0.004 & 1220$\pm$90  & 1730$\pm$30  & 5.66$\pm$0.06 & 6.04$\pm$0.02	\\
NGC\,4151\tablenotemark{f}  & 0.0026\tablenotemark{c} & 1988 RM\tablenotemark{d}	& \IUE/SWP    & 6.00 & 1465--1630 & 42.115$\pm$0.006 & 3590$\pm$100 & 4740$\pm$40  & 6.77$\pm$0.03 & 7.08$\pm$0.01	\\
 	   & 0.0026\tablenotemark{c} & 1991 RM\tablenotemark{d}	& \IUE/SWP    & 6.00 & 1465--1630 & 42.439$\pm$0.004 & 4890$\pm$110 & 4230$\pm$40  & 7.21$\pm$0.02 & 7.16$\pm$0.01	\\
 	   & 0.0026\tablenotemark{c} & 1998 Feb 10	& \HST/STIS\tablenotemark{e}& 1.20 & 1465--1630 & 42.708$\pm$0.001 & 3470$\pm$50  & 4480$\pm$10  & 7.06$\pm$0.02 & 7.35$\pm$0.01	\\
 	   & 0.0026\tablenotemark{c} & 1995 Mar 04,05	& HUT	     & 2.00 & 1465--1630 & 43.024$\pm$0.002 & 4720$\pm$90  & 3900$\pm$30  & 7.49$\pm$0.02 & 7.39$\pm$0.01	\\
 	   & 0.0026\tablenotemark{c} & 1995 Mar 07	& HUT	     & 2.00 & 1465--1630 & 43.057$\pm$0.003 & 3940$\pm$100 & 3450$\pm$50  & 7.35$\pm$0.02 & 7.31$\pm$0.02	\\
 	   & 0.0026\tablenotemark{c} & 1995 Mar 10	& HUT	     & 2.00 & 1465--1680 & 43.071$\pm$0.003 & 5090$\pm$140 & 5360$\pm$90  & 7.58$\pm$0.03 & 7.70$\pm$0.02	\\
 	   & 0.0026\tablenotemark{c} & 1995 Mar 13	& HUT	     & 2.00 & 1465--1680 & 43.084$\pm$0.003 & 4800$\pm$160 & 4760$\pm$80  & 7.54$\pm$0.03 & 7.60$\pm$0.02	\\
 	   & 	     &   	        & 	     & 	    & 	         & 	            & 	  	   & 	 & \boldmath$7.26\pm0.29$ & \boldmath$7.33\pm0.22$	\\
PG1226+032 & 0.15834 & 1991 Jan 14,15,17& \HST/FOS\tablenotemark{e} & 1.50 & 1720--1860 & 46.281$\pm$0.001 & 3470$\pm$80  & 3300$\pm$40  & 8.95$\pm$0.02 & 8.98$\pm$0.01	\\
 	   & 0.15834 & 1991 Dec 07, 12	& \IUE/SWP    & 6.00 & 1720--1860 & 46.308$\pm$0.005 & 4050$\pm$230 & 3230$\pm$140 & 9.10$\pm$0.05 & 8.97$\pm$0.04	\\
 	   & 0.15834 & 1992 Jan 05	& \IUE/SWP    & 6.00 & 1720--1860 & 46.301$\pm$0.007 & 3530$\pm$240 & 2860$\pm$160 & 8.98$\pm$0.06 & 8.86$\pm$0.05	\\
 	   & 0.15834 & 1992 Dec 17,28,29& \IUE/SWP    & 6.00 & 1720--1860 & 46.352$\pm$0.003 & 4040$\pm$170 & 3450$\pm$110 & 9.12$\pm$0.04 & 9.05$\pm$0.03	\\
 	   & 0.15834 & 1993 Jan 04-06,09& \IUE/SWP    & 6.00 & 1720--1860 & 46.355$\pm$0.004 & 3980$\pm$300 & 2570$\pm$170 & 9.11$\pm$0.07 & 8.80$\pm$0.06	\\
 	   & 0.15834 & 1993 Jan 16	& \IUE/SWP    & 6.00 & 1720--1860 & 46.263$\pm$0.006 & 3204$\pm$430 & 3750$\pm$190 & 8.87$\pm$0.12 & 9.08$\pm$0.05	\\
 	   & 0.15834 & 1994 May 15	& \IUE/SWP    & 6.00 & 1720--1860 & 46.311$\pm$0.006 & 3060$\pm$210 & 2840$\pm$190 & 8.86$\pm$0.06 & 8.86$\pm$0.06	\\
 	   & 	     &   	        & 	     & 	    & 	         & 	            & 	  	   & 	 & \boldmath$9.00\pm0.11$ & \boldmath$8.98\pm0.11$	\\
PG1229+204 & 0.06301 & 1982 May	        & \IUE/SWP    & 6.00 & 1583--1710 & 44.529$\pm$0.009 & 3410$\pm$240 & 2730$\pm$150 & 8.01$\pm$0.06 & 7.88$\pm$0.05	\\
 	   & 0.06301 & 1983 Jun	        & \IUE/SWP\tablenotemark{e} & 6.00 & 1583--1710 & 44.515$\pm$0.007 & 3640$\pm$210 & 2680$\pm$120 & 8.06$\pm$0.05 & 7.86$\pm$0.04	\\
 	   & 	     &   	        & 	     & 	    & 	         & 	            & 	  	   & 	 & \boldmath$8.04\pm0.05$ & \boldmath$7.87\pm0.04$\\
NGC\,4593\tablenotemark{f}  & 0.00865\tablenotemark{c}& 2002 Jun 24	& \HST/STIS\tablenotemark{e}& 0.51 & 1500--1615 & 42.575$\pm$0.007 & 2450$\pm$120 & 3000$\pm$50  & 6.68$\pm$0.04 & 6.93$\pm$0.02	\\
PG1307+085 & 0.15500 & 1993 Jul 21	& \HST/FOS\tablenotemark{e} & 2.20 & 1700--1880 & 44.941$\pm$0.005 & 3700$\pm$240 & 3380$\pm$90  & 8.30$\pm$0.06 & 8.29$\pm$0.03	\\
Mrk\,279\tablenotemark{f}   & 0.03045 & 2011 Jun 27	& \HST/COS\tablenotemark{e} & 0.62 & 1510--1675 & 43.057$\pm$0.004 & 4030$\pm$100 & 3180$\pm$30  & 7.37$\pm$0.02 & 7.24$\pm$0.01	\\
NGC\,5548\tablenotemark{f}  & 0.01717 & 1989 RM\tablenotemark{d}	& \IUE/SWP    & 6.00 & 1500--1655 & 43.594$\pm$0.009 & 4630$\pm$210 & 3860$\pm$90  & 7.78$\pm$0.04 & 7.69$\pm$0.02	\\
 	   & 0.01717 & 1993 RM\tablenotemark{d}	& \HST/FOS    & 1.90 & 1500--1655 & 43.485$\pm$0.001 & 3500$\pm$40  & 3920$\pm$10  & 7.48$\pm$0.01 & 7.64$\pm$0.01	\\
 	   & 0.01717 & 2011 Jun 16	& \HST/COS\tablenotemark{e} & 0.07 & 1480--1680 & 43.757$\pm$0.001 & 2710$\pm$50  & 5330$\pm$40  & 7.40$\pm$0.02 & 8.05$\pm$0.01	\\
 	   & 	     &   	        & 	     & 	    & 	         & 	            & 	  	   & 	 & \boldmath$7.47\pm0.20$ & \boldmath$7.80\pm0.23$	\\
PG1426+015 & 0.08647 & 1985 Mar 01,02	& \IUE/SWP\tablenotemark{e} & 6.00 & 1600--1755 & 45.180$\pm$0.004 & 4890$\pm$210 & 3760$\pm$130 & 8.66$\pm$0.04 & 8.51$\pm$0.03	\\
Mrk\,817   & 0.03145 & 2009 Aug 4	& \HST/COS\tablenotemark{e} & 0.21 & 1530--1665 & 44.318$\pm$0.001 & 4890$\pm$110 & 3280$\pm$20  & 8.21$\pm$0.02 & 7.93$\pm$0.01	\\
 	   & 0.03145 & 1981 Nov 6	& \IUE/SWP    & 6.00 & 1520--1670 & 44.051$\pm$0.011 & 4130$\pm$340 & 4820$\pm$150 & 7.92$\pm$0.07 & 8.12$\pm$0.03	\\
 	   & 0.03145 & 1981 Nov 7	& \IUE/SWP    & 6.00 & 1510--1690 & 44.016$\pm$0.007 & 4280$\pm$280 & 4910$\pm$130 & 7.93$\pm$0.06 & 8.12$\pm$0.03	\\
 	   & 0.03145 & 1982 Jul 18	& \IUE/SWP    & 6.00 & 1520--1690 & 44.115$\pm$0.005 & 4100$\pm$190 & 4530$\pm$110 & 7.95$\pm$0.04 & 8.10$\pm$0.02	\\
 	   & 	     &   	        & 	     & 	    & 	         & 	            & 	  	   & 	 & \boldmath$8.12\pm0.14$ & \boldmath$8.00\pm0.09$	\\
Mrk\,290   & 0.02958 & 2009 Oct 28	& \HST/COS\tablenotemark{e} & 0.21 & 1515--1680 & 43.581$\pm$0.001 & 1970$\pm$50  & 3720$\pm$20  & 7.03$\pm$0.02 & 7.65$\pm$0.01	\\
PG1613+658 & 0.12900 & 1990 Dec 02,05,10& \IUE/SWP    & 6.00 & 1690--1825 & 45.129$\pm$0.005 & 6250$\pm$300 & 3360$\pm$80  & 8.85$\pm$0.04 & 8.38$\pm$0.02	\\
 	   & 0.12900 & 2010 Apr 9	& \HST/COS\tablenotemark{e} & 0.30 & 1640--1860 & 45.318$\pm$0.002 & 5840$\pm$190 & 4840$\pm$50  & 8.89$\pm$0.03 & 8.80$\pm$0.01	\\
 	   & 	     &   	        & 	     & 	    & 	         & 	            & 	  	   & 	 & \boldmath$8.88\pm0.04$ & \boldmath$8.69\pm0.30$	\\
3C\,390.3  & 0.05610 & 1995,1996 RM\tablenotemark{d}	& \IUE/SWP    & 6.00 & 1580--1728 & 43.808$\pm$0.006 & 5840$\pm$150 & 4870$\pm$40  & 8.09$\pm$0.02 & 8.00$\pm$0.01	\\
 	   & 0.05610 & 1996 Mar 31	& \HST/FOS\tablenotemark{e} & 1.40 & 1530--1750 & 43.637$\pm$0.004 & 6120$\pm$240 & 5270$\pm$100 & 8.04$\pm$0.04 & 7.98$\pm$0.02	\\
 	   & 	     &   	        & 	     & 	    & 	         & 	            & 	  	   & 	 & \boldmath$8.07\pm0.04$ & \boldmath$8.00\pm0.02$	\\
Mrk\,509\tablenotemark{f}   & 0.03440 & 1992 Jun 22	& \IUE/SWP    & 6.00 & 1525--1670 & 44.402$\pm$0.010 & 5420$\pm$290 & 3410$\pm$130 & 8.34$\pm$0.05 & 8.01$\pm$0.03	\\
 	   & 0.03440 & 1992 Jun 21	& \HST/FOS    & 2.00 & 1525--1670 & 44.317$\pm$0.002 & 3940$\pm$150 & 4070$\pm$30  & 8.02$\pm$0.03 & 8.12$\pm$0.01	\\
 	   & 0.03440 & 1992 Oct 25,26,29&	\IUE/S& 6.00 & 1525--1670 & 44.593$\pm$0.007 & 4280$\pm$250 & 3710$\pm$120 & 8.24$\pm$0.05 & 8.18$\pm$0.03	\\
 	   & 0.03440 & 2009 Dec 10	& \HST/COS\tablenotemark{e} & 0.07 & 1535--1690 & 44.515$\pm$0.001 & 3220$\pm$40  & 3760$\pm$10  & 7.95$\pm$0.01 & 8.15$\pm$0.01	\\
 	   & 0.03440 & 2001 Apr 13	& \HST/STIS   & 0.42 & 1520--1680 & 44.250$\pm$0.003 & 3340$\pm$90  & 4240$\pm$90  & 7.84$\pm$0.03 & 8.12$\pm$0.02	\\
 	   & 	     &   	        & 	     & 	    & 	         & 	            & 	  	   & 	 & \boldmath$7.97\pm0.21$ & \boldmath$8.13\pm0.07$	\\
PG2130+099 & 0.06298 & 1995 July 24	& \HST/GHRS   & 0.65 & 1600--1687 & 44.517$\pm$0.003 & 2130$\pm$60  & 2230$\pm$40  & 7.59$\pm$0.03 & 7.70$\pm$0.02	\\
 	   & 0.06298 & 2010 Oct 28	& \HST/COS\tablenotemark{e} & 0.21 & 1580--1710 & 44.339$\pm$0.002 & 2250$\pm$40  & 2890$\pm$60  & 7.54$\pm$0.02 & 7.83$\pm$0.02	\\
 	   & 	     &   	        & 	     & 	    & 	         & 	            & 	  	   & 	 & \boldmath$7.56\pm0.04$ & \boldmath$7.76\pm0.09$	\\
NGC\,7469  & 0.01632 & 1996 RM\tablenotemark{d}	& \IUE/SWP    & 6.00 & 1510--1640 & 43.538$\pm$0.002 & 3120$\pm$90  & 3220$\pm$50  & 7.40$\pm$0.03 & 7.50$\pm$0.02	\\
 	   & 0.01632 & 1996 Jun 18	& \HST/FOS    & 1.40 & 1500--1650 & 43.428$\pm$0.002 & 2650$\pm$70  & 3310$\pm$40  & 7.20$\pm$0.03 & 7.47$\pm$0.01	\\
 	   & 0.01632 & 2010 Oct 16	& \HST/COS\tablenotemark{e} & 0.15 & 1520--1650 & 43.740$\pm$0.002 & 2800$\pm$90  & 2970$\pm$30  & 7.42$\pm$0.03 & 7.54$\pm$0.01	\\
 	   & 	     &   	        & 	     & 	    & 	         & 	            & 	  	   & 	 & \boldmath$7.33\pm0.12$ & \boldmath$7.50\pm0.04$	

\enddata

\tablenotetext{a}{The effective resolution we assume may be larger than
  the original, default instrumental resolution because we binned to a
  larger spectral dispersion in some cases, e.g., for COS spectra.}

\tablenotetext{b}{Values in bold are the uncertainty weighted mean of
  each object; see Section \ref{S_widthsMasses} for details.}

\tablenotetext{c}{This redshift has been modified to reflect the most
  probable true distance \citep[see][]{Bentz13} and is assumed for the
  calculation of the luminosity, assuming a cosmology with
  $\Omega_{m}=0.3$, $\Omega_{\Lambda}=0.70$, and $H_0 = 70$ km
  sec$^{-1}$ Mpc$^{-1}$.}

\tablenotetext{d}{Result is based on the mean reverberation mapping
  campaign spectrum for this object.  Original references for these
  campaigns: Fairall 9 \citep{Rodriguezpascual97}; NGC\,3783
  \citep{Reichert94}; NGC4151 \citep{Metzroth06}; NGC5548
  (\citealt{Clavel91}, \IUE; \citealt{Korista95}, \HST); 3C\,390.3
  \citep{OBrien98}; NGC\,7469 \citep{Wanders97}.}

\tablenotetext{e}{Spectrum shown in Figure \ref{fig:RMsamplefits}.}

\tablenotetext{f}{The \civ\ profile of this object was observed to have
  absorption across the line peak.}

\end{deluxetable}
\clearpage
\end{landscape}



\end{document}